%% file: main.tex
\documentclass[lettersize,journal]{IEEEtran}
\usepackage{amsmath,amsfonts}
\usepackage{algorithmic}
\usepackage{algorithm}
\usepackage{array}
\usepackage[caption=false,font=normalsize,labelfont=sf,textfont=sf]{subfig}
\usepackage{textcomp}
\usepackage{stfloats}
\usepackage{url}
\usepackage{verbatim}
\usepackage{graphicx}
\usepackage[nocompress]{cite}
\usepackage[hidelinks]{hyperref}
\usepackage{cleveref}

\usepackage{multicol}
\usepackage{multirow}
\usepackage{color, colortbl}
\usepackage[svgnames]{xcolor}

\usepackage{graphicx}
\usepackage{tabu}                      %
\usepackage{booktabs}                  %

\usepackage{mathptmx}                  %
\usepackage{subfig}
\usepackage{comment}
\usepackage{bm}
\usepackage{setspace}
\usepackage{soul}
\usepackage{enumitem}

\usepackage{colortbl}
\usepackage{rotating}

\usepackage{amssymb}   %
\usepackage{pifont}    %
\usepackage{xspace}
\usepackage{xcolor}
\usepackage{wrapfig}
\usepackage{caption}

\definecolor{NavyBlue}{RGB}{0,0,128}

\renewcommand{\paragraph}[1]{\vspace{4pt}\noindent\textbf{#1:}}
\Crefformat{figure}{#2Fig.~#1#3}
\Crefmultiformat{figure}{Figs.~#2#1#3}{ and~#2#1#3}{, #2#1#3}{ and~#2#1#3}
\Crefformat{section}{#2Sect.~#1#3}
\Crefmultiformat{section}{Sect.~#2#1#3}{ and~#2#1#3}{, #2#1#3}{ and~#2#1#3}

\newcommand{\lnorm}        {\ell}

\definecolor{sb20}{RGB}{198,219,239} %
\definecolor{sb40}{RGB}{123,175,212} %
\definecolor{sb70}{RGB}{33,113,181}  %
\definecolor{steelblue}{RGB}{70,130,180}

\newcommand{\alg}[1]{{\textsf{#1}}\xspace}

\newcommand{\methodGau}{\alg{Gaussian}}
\newcommand{\methodMean}{\alg{Mean}}
\newcommand{\methodMed}{\alg{Median}}
\newcommand{\methodMin}{\alg{Min}}
\newcommand{\methodMax}{\alg{Max}}
\newcommand{\methodSG}{\alg{Savitzky-Golay}}
\newcommand{\methodBW}{\alg{Butterworth}}
\newcommand{\methodCheb}{\alg{Chebyshev}}
\newcommand{\methodEllip}{\alg{Elliptical}}
\newcommand{\methodCut}{\alg{FFT Cutoff}}

\newcommand{\methodLTTB}{\alg{LTTB}}
\newcommand{\methodMFour}{\alg{M4}}
\newcommand{\methodMinMaxLTTB}{\alg{MinMaxLTTB}}
\newcommand{\methodUni}{\alg{Uniform Subsample}}
\newcommand{\methodDP}{\alg{Douglas-Peucker}}
\newcommand{\methodFPCS}{\alg{FPCS}}
\newcommand{\methodTDA}{\alg{TopoLines}}

\newcommand{\methodASAP}{\alg{ASAP}}
\newcommand{\methodPAA}{\alg{PAA}}

\begin{document}

\title{How to Simplify a Line Chart: A Taxonomy and Feature-Based Evaluation}

\author{Rifat Ara Proma, Paul Rosen

\thanks{Rifat Ara Proma is with the University of Utah. E-mail: rifat.proma@utah.edu.}
\thanks{Paul Rosen is with the University of Utah. E-mail: prosen@sci.utah.edu.}}

\markboth{Journal of \LaTeX\ Class Files,~Vol.~14, No.~8, August~2021}%
{Proma and Rosen: How to Simplify a Line Chart}

\maketitle

\begin{figure*}[!ht]
    \centering
    \captionsetup{labelfont=scriptsize,textfont=scriptsize}

    \includegraphics[width=\linewidth]{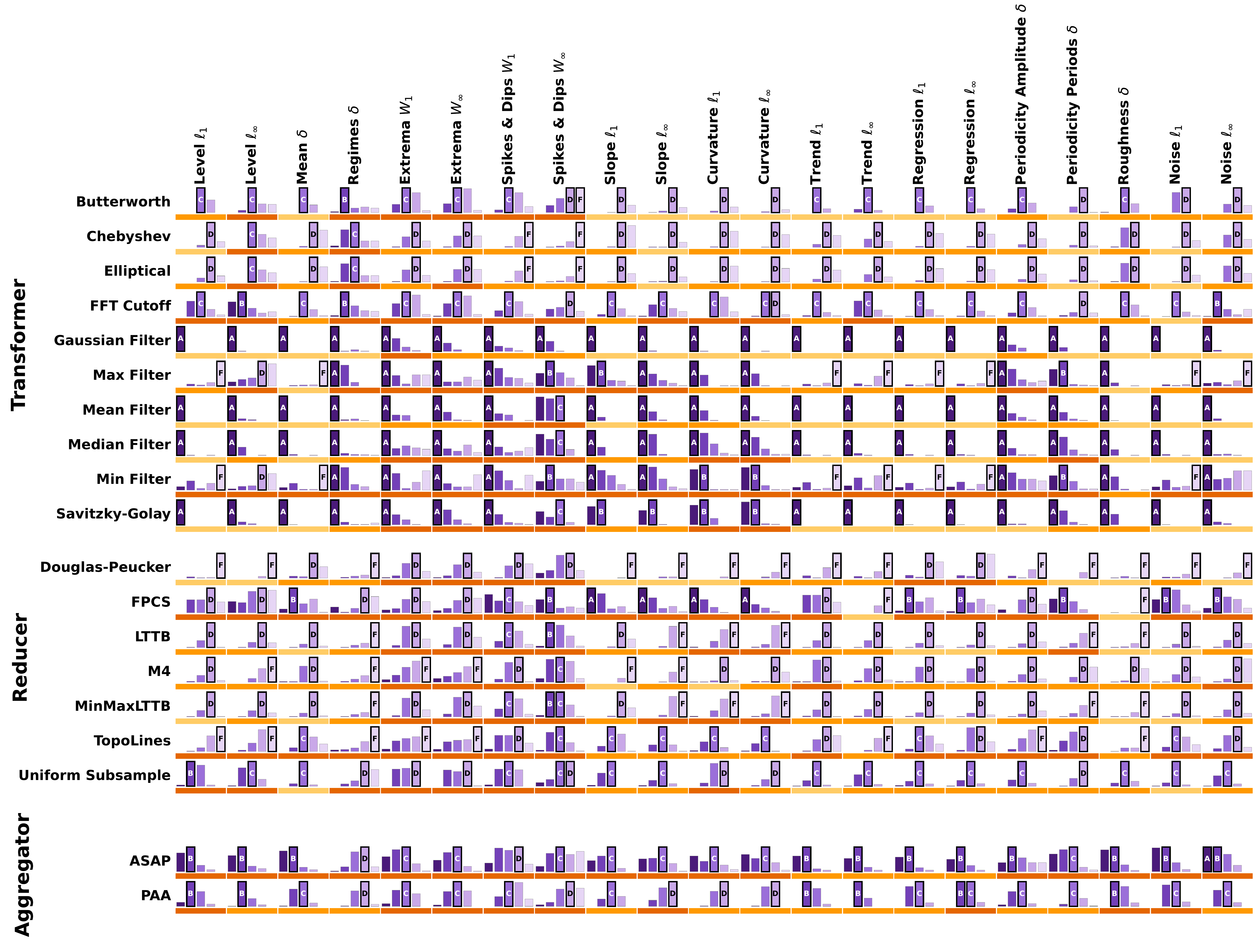}

    \vspace{4pt}

    \includegraphics[width=0.35\linewidth]{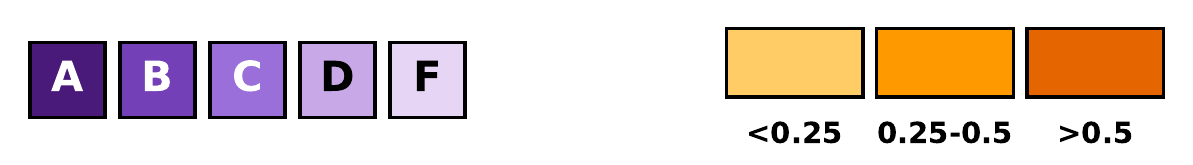}

   \caption{
    Performance evaluation of 19 simplification techniques across 21 feature preservation metrics. Each cell shows a grade distribution histogram (purple bars) aggregated across 80 datasets. The tallest bar indicates the \textit{mode grade}, the most frequently achieved grade across datasets (ties shown by highlighting multiple bars). The colored line beneath encodes \textit{deviation}, the mean absolute distance between each dataset's grade and the mode grade. 
    }
    \label{fig:teaser}
\end{figure*}

\input{sec.abstract}


\input{sec.intro}
\input{sec.prior_work}

\input{sec.line_chart_simplification_background}

\input{sec.experimental_setup}
\input{sec.feature_preservation}

\input{sec.evaluation}

\input{sec_results}

\input{sec.taxonomy}
\input{sec.discussion}

\input{sec.conclusion}

\setstretch{0.96}

\bibliographystyle{abbrv-doi-hyperref}

\bibliography{main}

\vspace{11pt}

\begin{IEEEbiography}[{\includegraphics[width=1in,height=1.25in,clip,keepaspectratio]{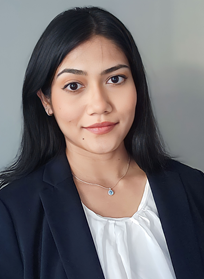}}]{Rifat Ara Proma}is a Ph.D. Candidate in the School of Computing at the University of Utah. Her dissertation focuses on developing strategies to mitigate visual clutter in line charts. She holds a Master's in Computer Science from the University of South Florida. Her research interests include information visualization, human-centric visualizations and human-computer interaction.
\end{IEEEbiography}

\begin{IEEEbiography}[{\includegraphics[width=1in,height=1.25in,clip,keepaspectratio]{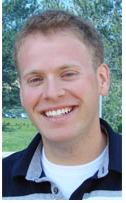}}]{Paul Rosen} is an Associate Professor at the University of Utah. He received his Ph.D.\ from Purdue University. His research interests include applying geometry- and topology-based approaches to problems in information visualization. Along with his collaborators, he has received best paper awards or honorable mentions at IEEE VIS, IEEE PacificVis, CG\&A, IVAPP, and SIBGRAPI.  Dr.\ Rosen received a National Science Foundation CAREER Award in 2019.
\end{IEEEbiography}

\vfill

\end{document}

%% file: sec.abstract.tex
\begin{abstract}
Simplifying large or noisy time-series is often a prerequisite for producing interpretable line charts, yet existing methods vary widely in how they alter \emph{visual features} such as trends, extrema, periodicity, curvature, roughness, and overall complexity. We introduce a feature-based taxonomy of line chart simplification that evaluates 19 commonly used algorithms across three families: Transformers, Reducers, and Aggregators. We also formalize 12 perceptually motivated, computationally defined visual features commonly used when analyzing time-series in line charts. We assess preservation using multiple similarity metrics per feature to capture different aspects of fidelity, yielding 21 distinct feature-preservation measurements. To capture the trade-off between fidelity and simplicity, we introduce a \emph{Feature--Complexity (FC)} score that combines feature preservation with complexity. Our systematic evaluation on 80 datasets reveals distinct family-level patterns: Transformers such as \methodGau, \methodMean, and \methodMed filters achieve A-mode grades on nearly all metrics but retain full temporal resolution; Reducers such as \methodDP and \methodMFour receive predominantly D-F mode grades trading fidelity for a reduction in the number of rendered points; and Aggregators \methodASAP and \methodPAA occupy a middle ground with primarily B-C mode grades. We further characterize the predictability of algorithm behavior across 
datasets: some methods produce consistent grades regardless of dataset type (\methodGau, \methodMean, \methodDP), while others exhibit high data-dependency, with performance varying substantially based on dataset characteristics (\methodFPCS, \methodASAP). An interactive website supporting exploration of all evaluated techniques, datasets, and results is available at \url{https://feature-taxonomy.vercel.app/}.
\end{abstract}

%% file: sec.intro.tex
\section{Introduction}
\IEEEPARstart{L}{ine} charts are among the most widely used techniques for visualizing time-series data because of their ability to convey essential characteristics such as trends, periodicity, extrema, and anomalies~\cite{few2008line, aigner2007visualizing}. Yet modern datasets often contain millions of samples, high-frequency noise, or complex multi-scale patterns that overwhelm displays and obscure important structure~\cite{chen2014data, hu2023application}. These challenges have motivated a wide variety of simplification methods, such as smoothing, sampling, filtering, and aggregation, each designed to reduce visual complexity while preserving relevant information.

Simplification, however, is not neutral. Smoothing can flatten peaks, sampling can discard structural extrema, aggregation can bias trends, and heavy compression can distort global statistics. Whether a method preserves or obscures the \emph{visual features} users rely on to make sense of the data shown in a line chart remains poorly understood. Although prior work has introduced topology-based~\cite{suh2019topolines}, value-preserving~\cite{van2023minmaxlttb}, feature-preserving~\cite{li2024fpcs}, and semantic-aware techniques~\cite{ahmed2022semantics}, there is no unifying framework that explains how these techniques differ in their effects on visual features or which methods are appropriate for analytic goals that require preserving specific visual features. Without such a structure, choosing a simplification method becomes ad hoc and difficult to justify.

This paper addresses this gap by introducing a feature-based taxonomy of line chart simplification. We evaluate 19 widely used simplification strategies across 80 diverse datasets, characterizing their ability to preserve 12 visual features: level, mean, regimes \& change-points, extrema, spikes \& dips, slope, curvature, trend, regression fit, periodicity, roughness, and noise. These features are motivated by prior studies of how viewers read and describe line charts, but we measure their preservation computationally rather than through human judgments. We leave the question of whether these computational measures agree with human perception to future work.

The taxonomy organizes simplification strategies into three conceptual families: transformers, reducers, and aggregators, based on their core operating principle on the signal or time-series (transforming values, selecting points, or combining both). Our evaluation characterizes feature preservation at the level of individual techniques and examines how strategies associated with each operating principle behave across datasets.

Importantly, the taxonomy is used as an organizational and generalization scaffold rather than as a predictor of within-family behavior. We do not assume uniform performance among methods in the same family. Instead, family labels are used to 
(i)~structure comparisons across a large design space, 
(ii)~support reasoning about unevaluated techniques with similar underlying mechanisms, and 
(iii)~motivate hybrid designs that intentionally combine transformation and reduction.

To compare simplification technique behavior across different types of data and features, we introduce a Feature-Complexity (FC) score, a unified measure of how well a simplification preserves a visual feature relative to the complexity it removes. Building on FC scores, our evaluation framework summarizes the performance of each simplification technique across its parameter range and assesses how consistently it behaves across different datasets. Together, these evaluations reveal how each simplification technique affects specific features, and which strategies are stable versus highly sensitive to dataset characteristics.

Our contributions are:
\begin{itemize}
    \item A feature-based taxonomy that organizes simplification techniques into transformers, reducers, and aggregators, evaluates multiple methods from each family in terms of preserving 12 visual features (level, mean, regimes \& change-points, extrema, spikes \& dips, slope, curvature, trend, regression fit, periodicity, roughness, and noise).

    \item A unified evaluation framework combining Feature-Complexity (FC) scoring, quartile-based grading, and deviation analysis to quantify feature preservation across 19 simplification techniques, 100 simplification degrees, and 80 real-world datasets.
    
    \item A systematic characterization of simplification technique behavior, identifying for each feature the simplification techniques that are consistently strong, consistently weak, or strongly data-dependent, and revealing global performance patterns across simplification technique families.

\end{itemize}

This taxonomy and evaluation provide a foundation for feature-aware and data-aware simplification, clarifying the trade-offs inherent in smoothing, sampling, and aggregation, and enabling more deliberate choices in the design of simplified line charts.

%% file: sec.prior_work.tex
\section{Prior Work}

\begin{figure}[!ht]
    \centering
    \captionsetup{labelfont=scriptsize,textfont=scriptsize}

    \newcommand{\quadw}{0.48\linewidth}
    \newcommand{\vsep}{8pt}
    \newcommand{\stacksep}{6pt}

    \begin{minipage}[t]{\quadw}
        \centering
        \vspace{20pt}
         {\scriptsize Input EEG signal}
        \includegraphics[width=\linewidth]{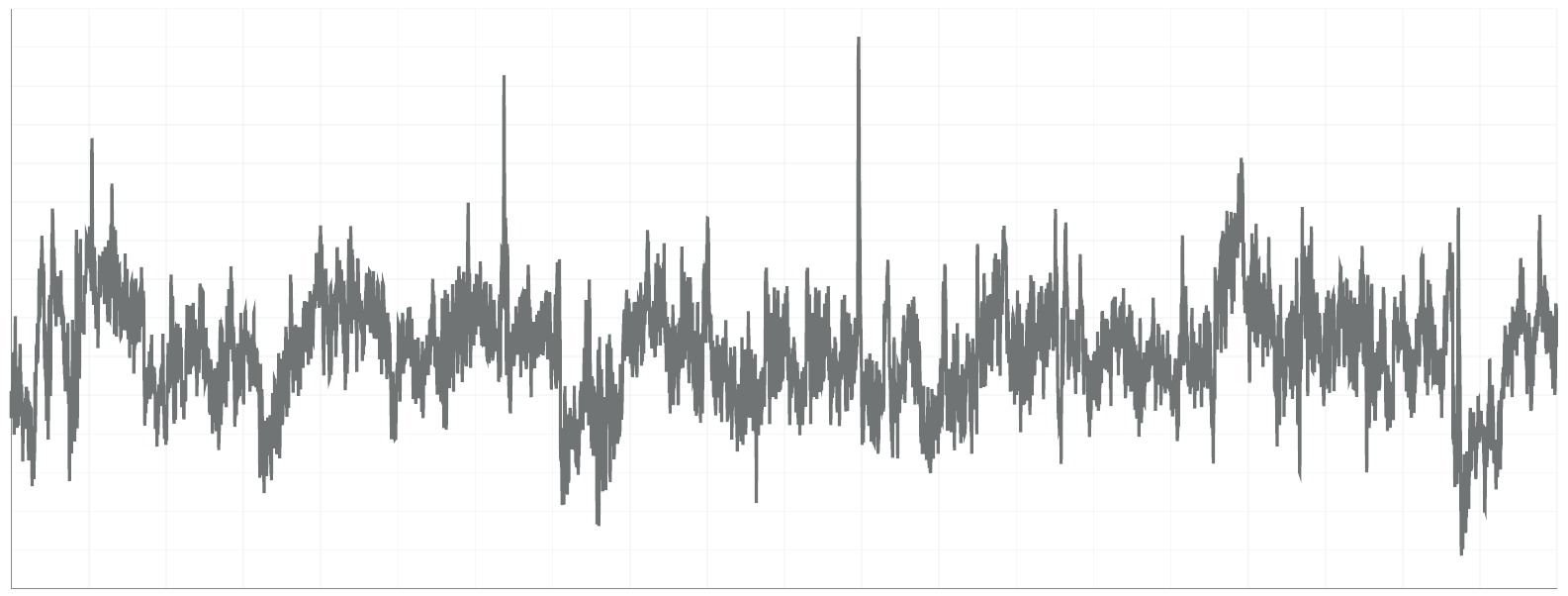}\\[-2pt]       
        \vfill
    \end{minipage}
    \hfill
    \begin{minipage}[t]{\quadw}
        \centering
        {\scriptsize Transformers}\\[\vsep]
        \includegraphics[width=\linewidth]{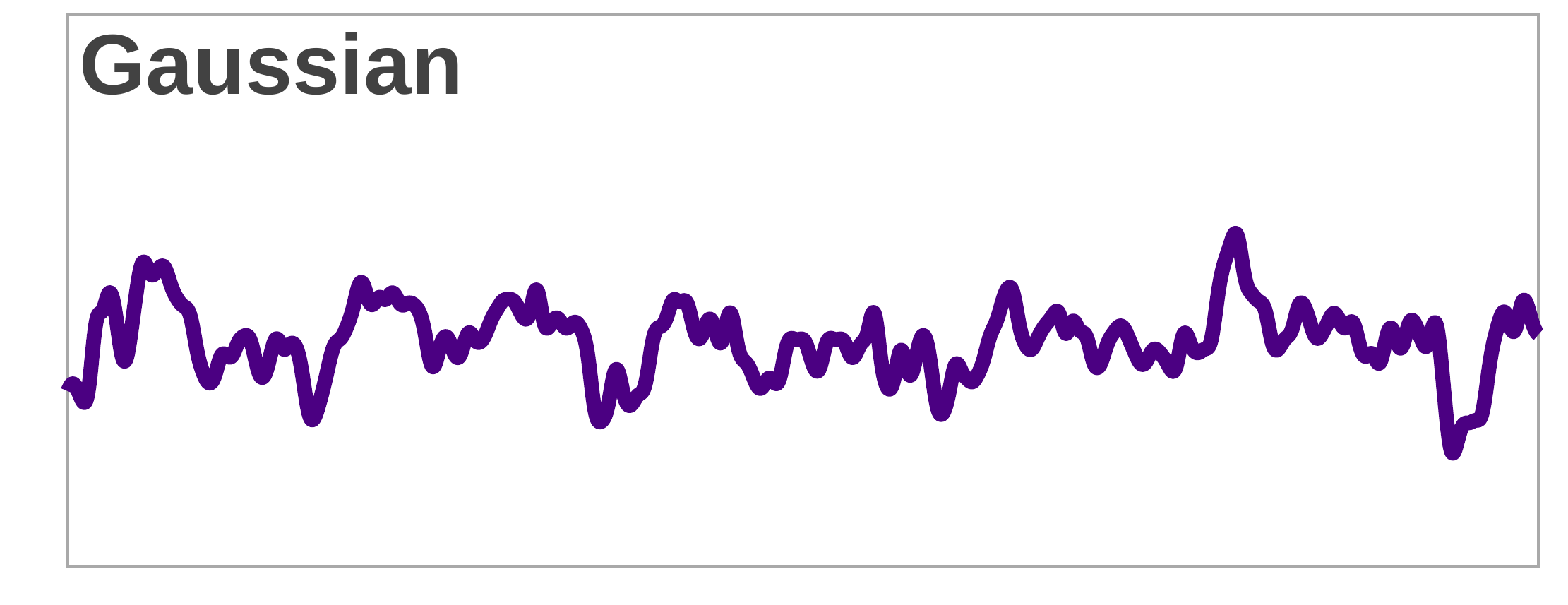}\\[\stacksep]
        \includegraphics[width=\linewidth]{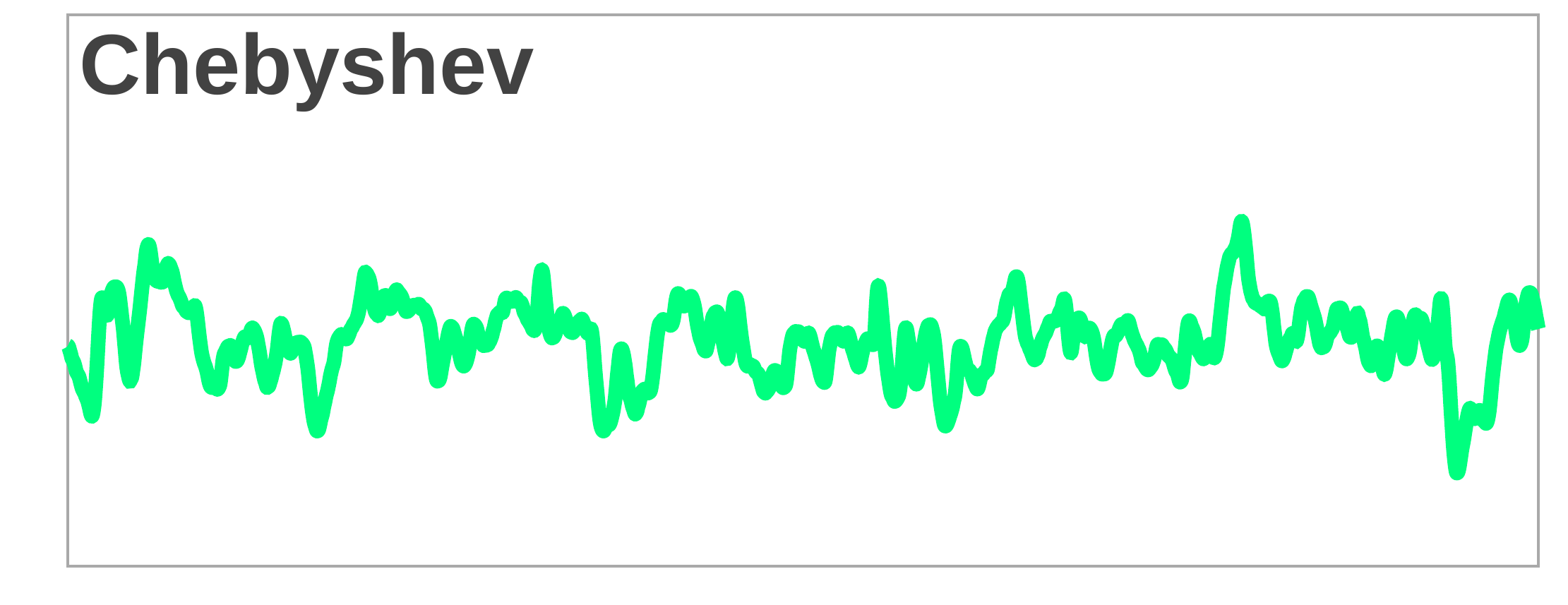}
    \end{minipage}

    \vspace{10pt}

    \begin{minipage}[t]{\quadw}
        \centering
        {\scriptsize Reducers}\\[\vsep]
        \includegraphics[width=\linewidth]{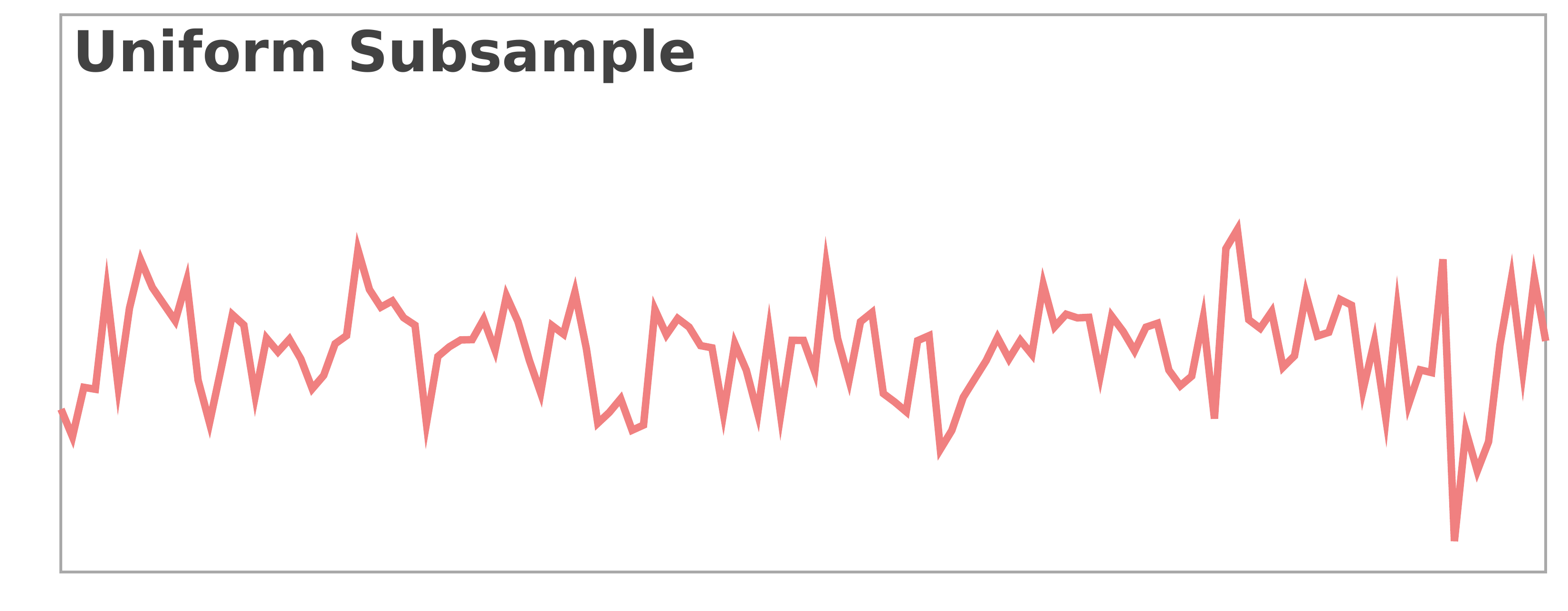}\\[\stacksep]
        \includegraphics[width=\linewidth]{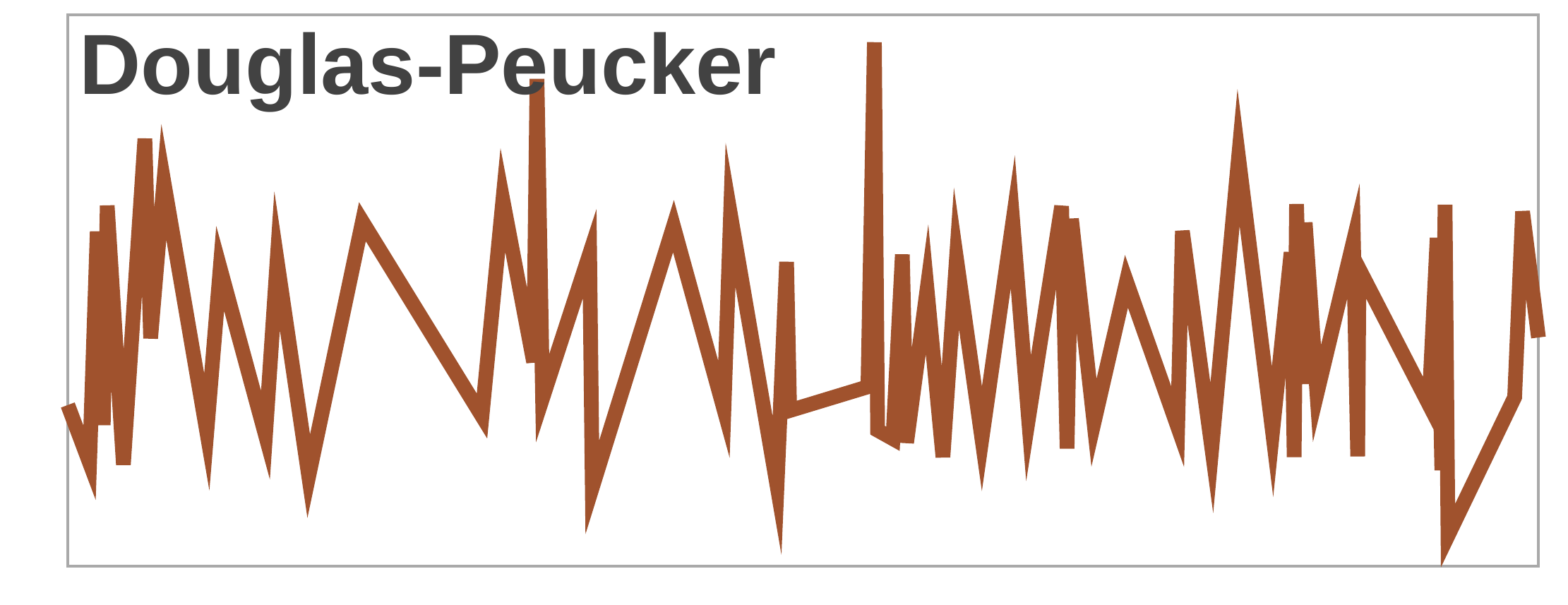}
    \end{minipage}
    \hfill
    \begin{minipage}[t]{\quadw}
        \centering
        {\scriptsize Aggregators}\\[\vsep]
        \includegraphics[width=\linewidth]{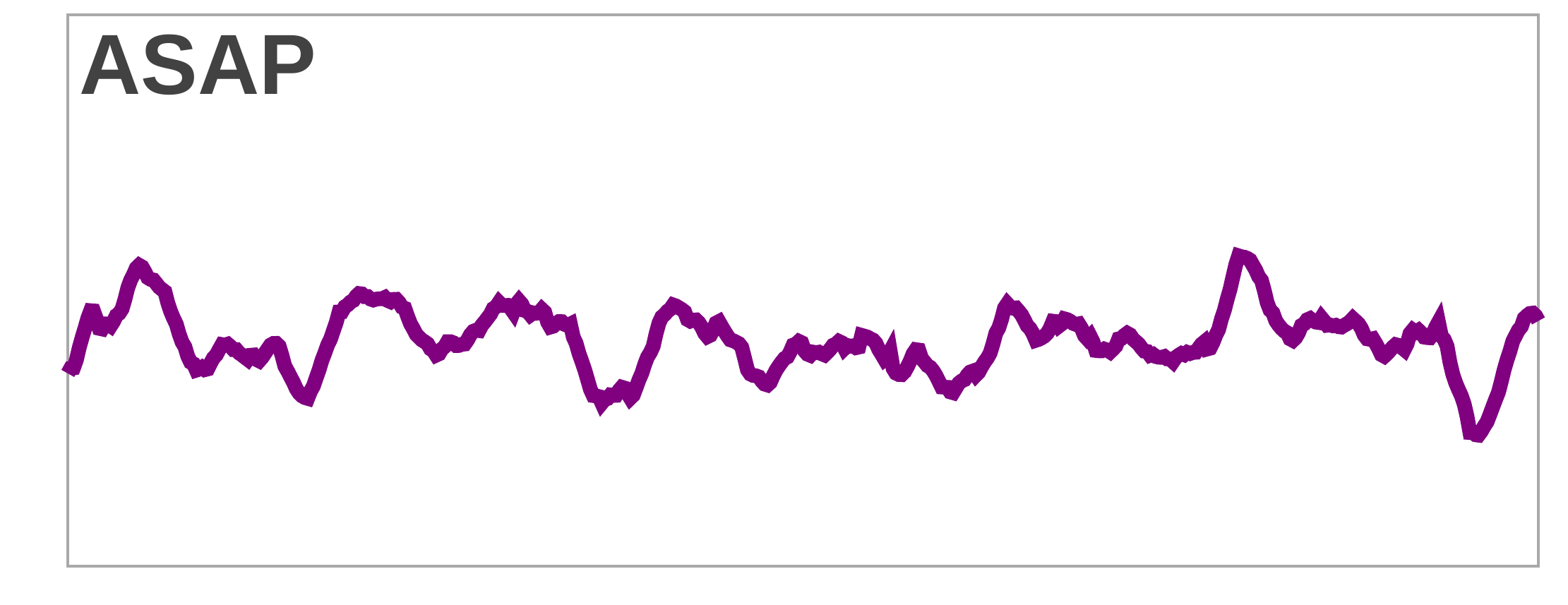}\\[\stacksep]
        \includegraphics[width=\linewidth]{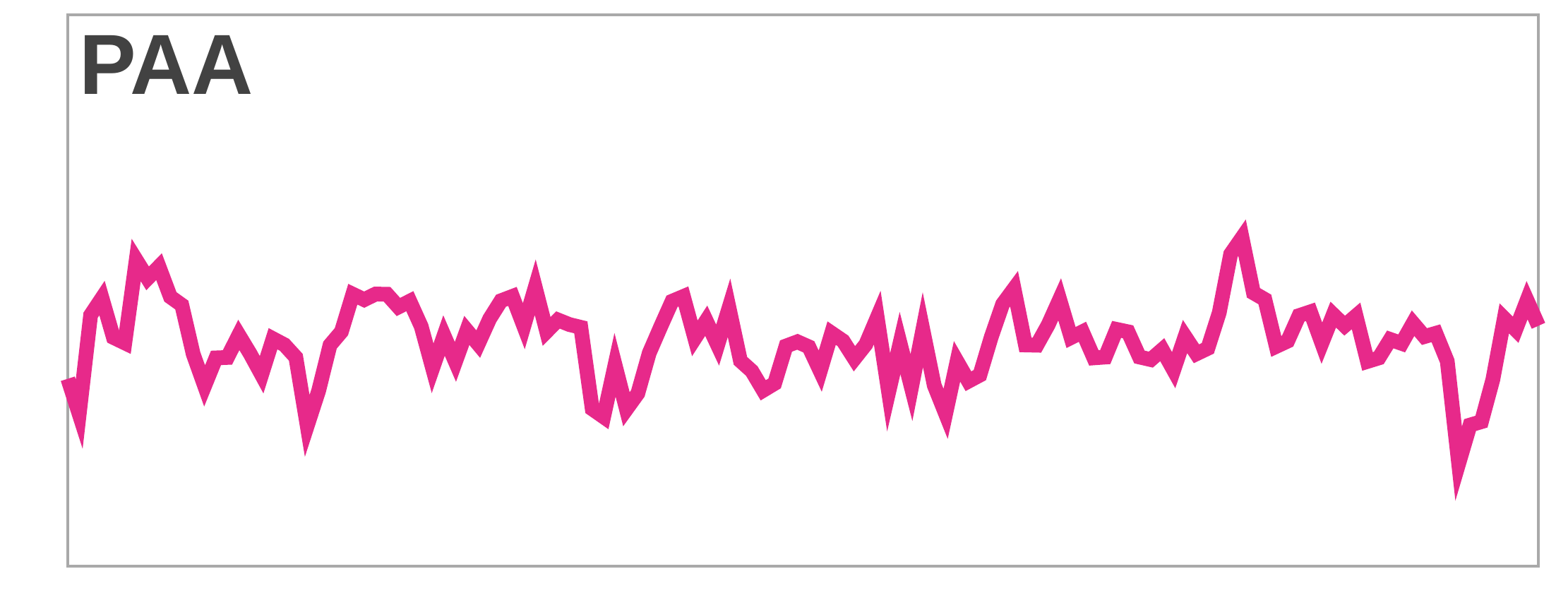}
    \end{minipage}

    \caption{Example outputs from three simplification families applied to an EEG signal dataset \texttt{eeg\_05\_10000} (10,000 samples, original PAE=1.539). \textbf{Transformers} modify values at all time steps without changing sample positions: \methodGau ($\sigma=41.4$, PAE=0.412) and \methodCheb filter (cutoff=0.010 Hz, PAE=0.604). \textbf{Reducers} select a subset of original samples: \methodUni (131 output points, PAE=0.556) and  \methodDP (67 output points, PAE=0.565). \textbf{Aggregators} combine adjacent data points into summary values: ASAP (resolution=505, PAE=0.341) and PAA (130 bins, PAE=0.461). All methods substantially reduce visual complexity (PAE reduced by 60--78\% from original), yet produce qualitatively different visual structures.}
    \label{fig:simplification_examples}
\end{figure}

We organize prior work into two areas: simplification strategies proposed for line charts, and studies of perceptual properties and visual features that shape how line charts are interpreted.

\subsection{Simplification for Time-Series Visualization}

A significant body of work examines methods for reducing visual complexity in time-series visualizations. Taxonomies of clutter reduction, such as Ellis and Dix~\cite{ellis2007taxonomy}, classify techniques that include sampling, filtering, and display-level adjustments, such as opacity and point size. Peng et al.~\cite{peng2004clutter} surveyed analogous strategies for multidimensional visualization. These works focus on reducing visual density and improving readability, but do not analyze how simplification alters the perceptual structures conveyed by the data.

To determine which strategy to use for reducing visual complexity, prior work has evaluated multiple sampling, smoothing, and aggregating techniques against a range of metrics. Van der Donckt et al.~\cite{van2023data} evaluate four downsampling techniques (EveryNth, MinMax, \methodLTTB~\cite{steinarsson2013downsampling}, \methodMFour~\cite{jugel2014m4}) for visualizing large-scale data on line charts, using image-based similarity metrics to quantify stability under zooming and panning. Their analysis also focuses on how well a downsampled time-series line chart visually approximates the original data using image quality assessment measures. They found that MinMax and \methodLTTB tend to preserve visual approximation fairly well, which motivated the development of \methodMinMaxLTTB~\cite{van2023minmaxlttb}, a method that combines the strengths of the two approaches. However, their work focuses primarily on sampling strategies and overall image-based similarity, rather than on how well individual visual features such as trend, extrema, and periodicity are preserved.

Rong and Bailis~\cite{rong2017asap} introduced a smoothing-based aggregation method, \methodASAP, and compared its performance with that of \methodMFour, \methodPAA~\cite{keogh2001dimensionality}, and the Visvalingam-Whyatt algorithm~\cite {visvalingam2017line}. They evaluated pixel accuracy and conducted a user study on anomaly detection. Their findings showed that \methodMFour was most effective at preserving pixel-level accuracy, while \methodASAP enabled users to detect anomalies more reliably. They therefore recommended combining these methods to achieve the benefits of both. This motivates the importance of understanding the strengths and weaknesses of individual algorithms, as complementary techniques can often be paired effectively. Our taxonomy is designed to support this type of informed pairing.

Taken together, prior work provides valuable insights into sampling efficacy, smoothing behavior, and general clutter-reduction strategies. However, much of this literature evaluates methods through task-based performance. LineSmooth~\cite{rosen2020linesmooth}, for example, compares nine smoothing algorithms across eight low-level analytical tasks, including retrieving values, computing derived values, and identifying anomalies, providing one of the most detailed evaluations of the effects of smoothing on task performance. Yet, as Quadri et al.~\cite{quadri2024you} noted, there is often a gap between what visualization designers intend viewers to see and what viewers actually perceive. Most design practices utilize tasks to create visualizations~\cite{amar2005low}, but such evaluations alone do not fully capture whether a visualization effectively emphasizes the key patterns in the data~\cite{quadri2021survey}. Building on this, our study evaluates simplification strategies at the level of individual visual features, rather than through aggregate task performance or overall image similarity. To the best of our knowledge, no prior work has taken this approach.

\subsection{Visual Feature Preservation for Line Chart Perception}
Prior work has shown that viewers primarily make sense of line charts through visual features rather than individual data values. Studies on how viewers describe and annotate line charts highlighted viewers looking for peaks and valleys, trend segments, and notable changes as the units of interpretation. Bromley and Setlur~\cite{bromley2023difference} analyzed how people distinguish and describe visual features, showing that descriptions focus on the height, shape, and prominence of local extrema. Kong and Agrawala~\cite{kong2009perceptual} demonstrate that common annotations on line charts map onto semantic concepts such as extrema, ranges, and trends. Xi~\cite{xi2010aspects} and Kim et al.~\cite{kim2021towards} similarly report that natural-language summaries of line charts emphasize patterns of trends, local peaks, and important changes over time.

Other studies have focused on which visual features viewers attend to and which structures they prioritize. Eye-tracking studies of dashboards and line graphs show that viewer attention is not uniform while looking at a line chart; rather, they pay more attention to certain areas and features~\cite{orlov2016eye}. A study by Proma et al.~\cite{proma2025visual} revealed that viewers attend to different visual features differently while looking at a line chart. The study reported that viewers consistently prioritized trends, but peaks and valleys were often ignored in noisy line charts. Systems such as Line Graph Explorer~\cite{kincaid2006line} and accessibility tools like Grapho~\cite{singh2022grapho} explicitly encode or surface features, including extrema, abrupt changes, and monotonic segments, to support navigation and explanation of complex line charts. Together, these results reinforce the idea that features like trends, extrema, and changes in behavior are the primary carriers of meaning in time-series visualizations.

Despite the importance of these perceptual structures, existing work lacks a systematic study of how different simplification strategies impact the preservation of such features. Prior studies focused on task performance rather than feature-level effects. In this paper, we address this gap by evaluating a wide range of commonly used simplification strategies and quantifying their impact on these visual features using computational preservation metrics across diverse real-world datasets.

%% file: sec.line_chart_simplification_background.tex
\section{Background on Line Chart Simplification}
Line chart simplification encompasses methods for reducing visual complexity while preserving salient data features. Building on the strategies reviewed in the prior section, we now establish the specific techniques and visual features that our evaluation relies on. We review two strands of background: (1)~different types of simplification techniques, and (2)~visual features relevant to the interpretation of line charts. The techniques are the methods we evaluate, and the features are the properties whose preservation we measure. Our evaluation quantifies how each technique preserves each feature. These definitions provide the shared vocabulary used throughout the paper and form the basis for our taxonomy and evaluation.

\input{sub_sec.data_simplification}

\input{sub_sec.visual_feratures}

%% file: sub_sec.data_simplification.tex
\subsection{Data Simplification Methods}

\begin{figure*}[!t]
\centering
    \subfloat[{\scriptsize \emph{Data Transformer}: \methodMean Filter}\label{fig:data_transformation_illustration}]{\includegraphics[width=0.25\linewidth]{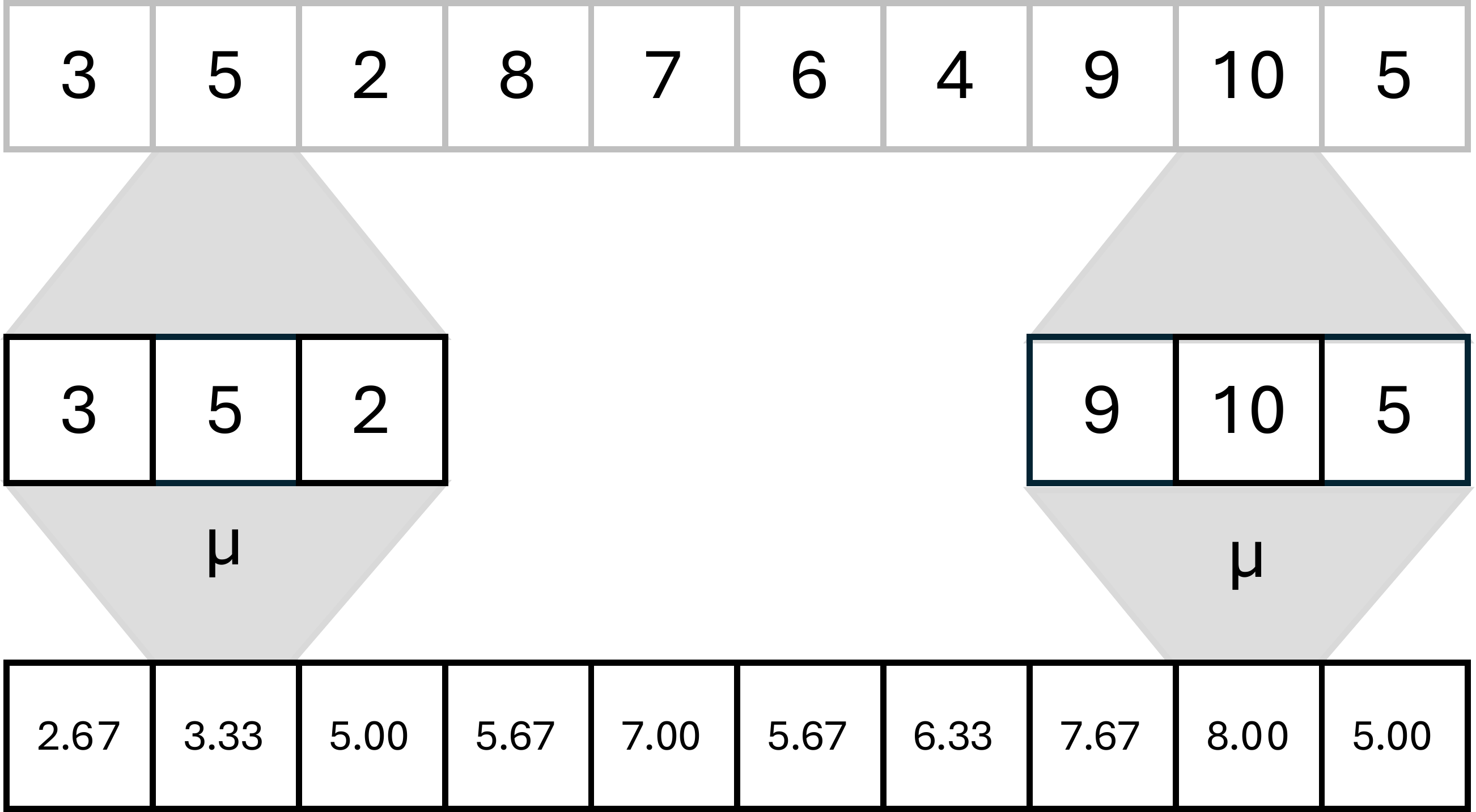}}
    \hspace{15pt}
    \subfloat[{\scriptsize \emph{Data Reducer}: \methodUni}\label{fig:data_reduction_illustration}]{\includegraphics[width=0.25\linewidth]{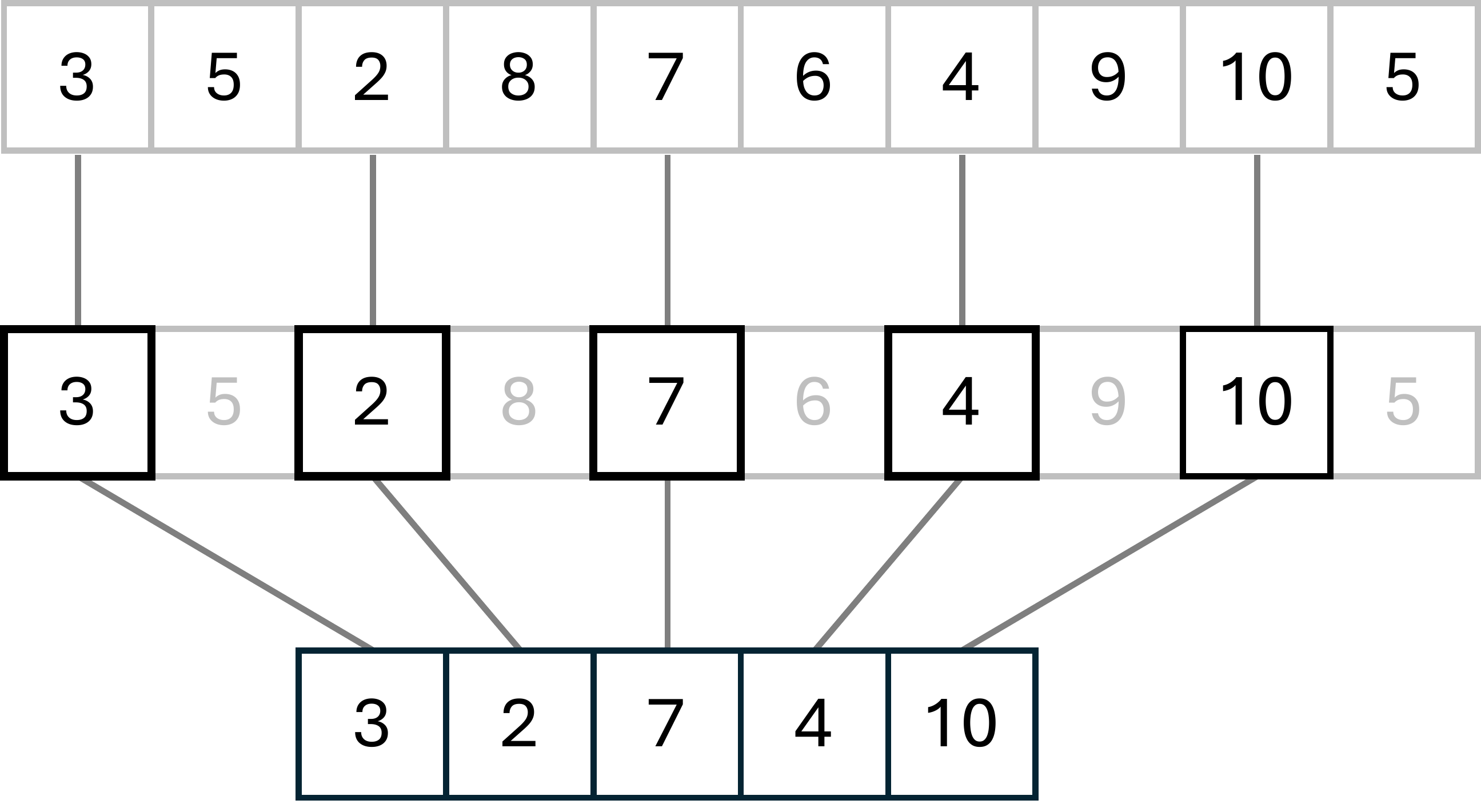}}
     \hspace{15pt}
    \subfloat[{\scriptsize \emph{Data Aggregator}: \methodPAA aggregation}\label{fig:data_aggregation_illustration}]{\includegraphics[width=0.25\linewidth]{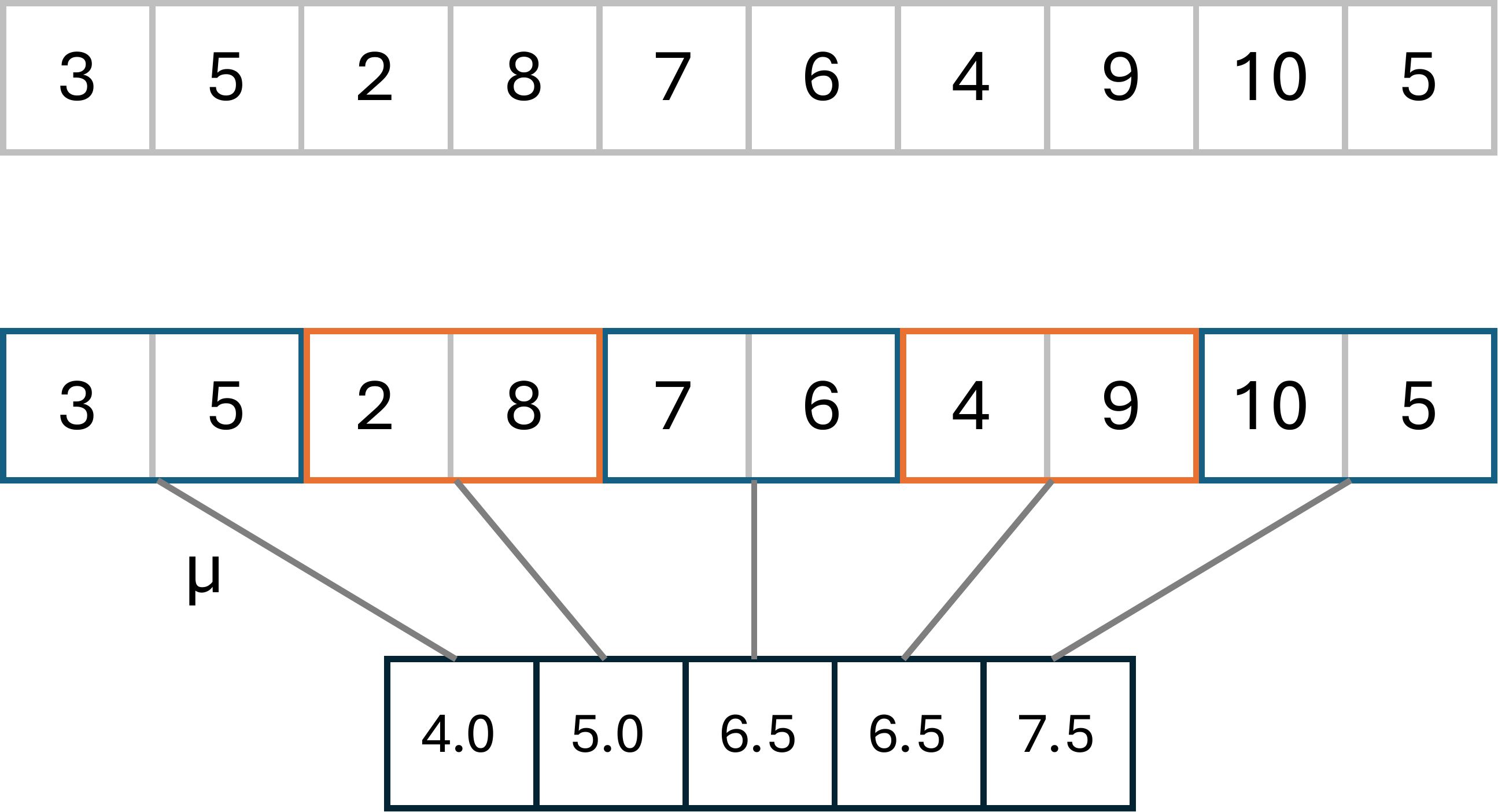}}
     \hspace{15pt}
    \caption{Illustration of the three categories of data simplification techniques : (a)~\emph{Data Transformer}: \methodMean Filter replaces each point with the local average over a sliding window; (b)~\emph{Data Reducer}: \methodUni selects a fixed subset of points at regular intervals; and  (c)~\emph{Data Aggregator}: \methodPAA partitions the domain into bins and replaces each bin with its mean.}
    \label{fig:preprocessing}
\end{figure*}

Data simplification refers to the process of making large and complex data more usable and interpretable~\cite{berman2016data}. Data simplification for line charts encompasses a wide range of techniques that differ not only in how they simplify the signal, but also in which aspects of the underlying data they aim to preserve. Some approaches employ straightforward subsampling strategies, such as \methodUni, while more sophisticated point–selection methods identify topologically significant samples~\cite{suh2019topolines} or Feature–Preserving Compensated Sampling (\methodFPCS) for streaming data~\cite{li2024fpcs}. Other methods operate in pixel space, selecting points based on how the line will be rendered on screen, such as M4~\cite{jugel2014m4} or \methodMinMaxLTTB~\cite{van2023minmaxlttb}. A large family of techniques applies numeric filters such as \methodGau~\cite{kopparapu2011identifying}, \methodMed, or \methodSG~\cite{savitzky1964smoothing} smoothing, while algorithms like \methodASAP~\cite{rong2017asap} prioritize producing an aggressively smoothed curve subject to statistical criteria.

Although these techniques differ in their mechanisms and in how they balance visual fidelity, statistical fidelity, and perceptual relevance, we categorize the algorithms studied in this paper into three families based on their core operating principle on the signal: \emph{data transformers}, \emph{data reducers}, and \emph{data aggregators}. These families are defined by how each algorithm operates on the input time-series, whether it modifies all data values, selects a subset of the original values, or combines multiple values into aggregated representations. This categorization provides a structured way to organize and compare technique-level results, and to reason about the behavior of methods that operate on the signal in similar ways.

\begin{table*}[!h]
\centering
\caption{Overview of transformer, reducer, and aggregator methods with examples, their main parameter, and 
computational complexity. 
Here, $n$ is the number of input points; $w$ is the window size; $\sigma$ is the standard deviation of a Gaussian 
kernel; $s$ is the target output size; and $c$ is the number of local minima or maxima.}

\label{tab:data_simplification_categories}
\resizebox{\textwidth}{!}{
\begin{tabular}{p{0.17\linewidth} p{0.48\linewidth} p{0.36\linewidth} p{0.31\linewidth} p{0.2\linewidth}}
\toprule
\textbf{Method} & \textbf{Description} & \textbf{Examples} & \textbf{Main Parameter} & \textbf{Complexity} \\
\midrule

Transformer &
Keeps all time steps but replaces each value with a smoothed or filtered estimate &
\begin{tabular}[t]{@{}l@{}}
Gaussian filter \\ 
Mean filter \\ 
Median filter \\ 
Min filter \\ 
Max filter \\
Savitzky-Golay \\ 
FFT cutoff filter \\ 
Butterworth \\ 
Chebyshev \\ 
Elliptical
\end{tabular} &
\begin{tabular}[t]{@{}l@{}}
standard deviation \\ 
window length \\ 
window length \\ 
window length \\ 
window length \\ 
poly window length \\ 
cutoff frequency \\ 
cutoff frequency \\ 
cutoff frequency \\ 
cutoff frequency
\end{tabular} &
\begin{tabular}[t]{@{}l@{}}
$O(n \cdot \sigma)$ \\ 
$O(n)$ \\ 
$O(n \log w)$ \\ 
$O(n)$ \\ 
$O(n)$ \\ 
$O(n \cdot w)$ \\ 
$O(n \log n)$ \\ 
$O(n \log n)$ \\ 
$O(n \log n)$ \\ 
$O(n \log n)$
\end{tabular} \\

\midrule
Reducer &
Selects a subset of original points and discards the rest while keeping exact values of retained points &
\begin{tabular}[t]{@{}l@{}}
Uniform subsampling \\ 
Douglas-Peucker (RDP) \\ 
LTTB \\ 
MinMaxLTTB \\ 
M4 \\ 
FPCS \\ 
TopoLines
\end{tabular} &
\begin{tabular}[t]{@{}l@{}}
sampling interval\\
target size\\
target size\\
target size\\
target size\\
sampling rate\\
persistence threshold
\end{tabular} &
\begin{tabular}[t]{@{}l@{}}
$O(n + s)$ \\ 
$O(s \log n)$ \\ 
$O(n)$ \\ 
$O(n)$ \\ 
$O(n)$ \\ 
$O(n)$ \\ 
$O(n + c \log c)$
\end{tabular} \\

\midrule

Aggregator &
Reduces and transforms by summarizing groups of points into new representative values &
\begin{tabular}[t]{@{}l@{}}
ASAP \\ 
PAA 
\end{tabular} &
\begin{tabular}[t]{@{}l@{}}
target resolution \\ 
number of bins
\end{tabular} &
\begin{tabular}[t]{@{}l@{}}
$O(n \log n)$ \\ 
$O(n)$
\end{tabular} \\

\bottomrule
\end{tabular}}
\end{table*}

\subsubsection{Data Transformer}
Transformers retain the same number of data points as the input time-series but replace each value with a modified estimate, typically by incorporating information from a local neighborhood or through a frequency–based transformation as illustrated in ~\Cref{fig:data_transformation_illustration}. These methods smooth or reshape the time-series without removing points. 

In this study, we considered ten data transformer algorithms listed in~\Cref{tab:data_simplification_categories}. Among these, the \methodGau filter applies a normal distribution kernel to emphasize nearby points and suppress high-frequency variation. The degree of simplification is controlled by the kernel's standard deviation $\sigma$, which determines how broadly the weights spread across the neighborhood. A family of sliding–window filters replace each point with a statistic computed from its local neighborhood. The \methodMean{} filter assigns the arithmetic mean of the window, the \methodMed{} filter selects the median, and the \methodMin{} and \methodMax{} filters take the minimum or maximum value within the window, respectively. \methodSG{}~\cite{savitzky1964smoothing} performs local polynomial regression within each window and replaces the center point with the value of the fitted polynomial. In all cases, a window of fixed length moves across the series, with larger windows introducing more substantial transformation of the data. 

The \methodCut{} filter operates by transforming the input series into the frequency domain using Fast Fourier Transformation (FFT), zeroing out all frequency components above a user–specified cutoff, and then applying the inverse FFT to reconstruct a simplified signal~\cite{cooley1965algorithm}. In addition to this cutoff operation, we also include three common infinite-impulse-response (IIR) low-pass filters such as \methodBW{}~\cite{butterworth1930theory}, \methodCheb{}~\cite{rhodes1980generalized}, and \methodEllip{}~\cite{cauer1958synthesis}. These filters differ in how sharply they transition from passband to stopband, with \methodBW being smoothest, \methodCheb steeper with controlled ripple, and \methodEllip offering the steepest cutoff with ripple in both bands. The passband contains the frequencies that the filter preserves, while the stopband contains the frequencies it attenuates or removes.

\subsubsection{Data Reducer}
Reducers simplify a time-series by selecting a subset of its original data points and discarding the rest, thereby lowering complexity without modifying the retained values. Unlike transformers, which operate on all samples, reducers return a shorter sequence whose points are drawn directly from the input. The final simplified chart is obtained by connecting these selected points, as illustrated in ~\Cref{fig:data_reduction_illustration}.

We considered seven reducer algorithms listed in~\Cref{tab:data_simplification_categories}. The \methodUni{} method keeps every $k$-th point, distributing samples evenly across the domain. \methodDP{}~\cite{douglas1973algorithms} starts from the endpoints of the polyline and iteratively refines it by repeatedly splitting the segment with the largest deviation from the straight line between its endpoints, using a priority queue, until a target number of points is reached. Shape-based reducers such as \methodLTTB{}~\cite{steinarsson2013downsampling} preserve large-scale structure by selecting points that maximize triangle areas within fixed bins. The hybrid \methodMinMaxLTTB{}~\cite{van2023minmaxlttb} enhances the efficiency of \methodLTTB{} by first identifying coarse-scale minima and maxima through a preselection, then applying LTTB to that reduced set of candidates. \methodMFour{}~\cite{jugel2014m4} aggregates each bin into four points (first, last, minimum, and maximum), ensuring that the simplified line rasterizes identically to the full-resolution series while achieving high reduction rates. All of these methods determine the level of simplification primarily through the chosen output size.

Two reducers in our selection explicitly target feature-bearing points rather than bins or geometric deviation. \methodFPCS{}~\cite{li2024fpcs} operates in a streaming manner, examining the incoming series at a user–controlled rate and emitting points only when they represent strong local changes. Lower rates retain a denser set of feature points, while higher rates yield a more aggressive reduction. \methodTDA{}~\cite{suh2019topolines} uses topological persistence to identify structurally significant extrema and simplifies the series by retaining only those points whose prominence exceeds a chosen threshold, thereby discarding minor fluctuations while keeping the most robust peaks and valleys.

\subsubsection{Data Aggregator}
Aggregators simplify a time-series by partitioning the domain into contiguous groups and replacing each group with a representative statistic, thereby combining aspects of both transformation and reduction. The resulting simplified chart is obtained by connecting these aggregated values, as illustrated in~\Cref{fig:data_aggregation_illustration}. 

We considered two aggregation methods listed in~\Cref{tab:data_simplification_categories}. The first is Piecewise Aggregate Approximation (\methodPAA{})~\cite{keogh2001dimensionality}, in which the time-series is divided into fixed-width bins and each bin is represented by the mean of its contained samples. The degree of simplification is determined by the number and width of bins. The second method is \methodASAP{}~\cite{rong2017asap}, which automatically selects an aggregation level that balances noise suppression with the preservation of large-scale structure. \methodASAP{} adapts its granularity based on the data, producing coarser aggregates in stable regions and finer detail where the signal exhibits meaningful deviations. The level of simplification is determined by a resolution parameter that specifies the target number of output points: lower resolutions yield heavier aggregation, whereas higher resolutions preserve more detail.

%% file: sub_sec.visual_feratures.tex
\subsection{Visual Features in Line Charts}
\label{sec:features}
When discussing \textit{visual features} in line charts, we refer not to design elements such as line thickness, color, or gridlines, but to data-driven characteristics that manifest visually and shape interpretation. These features capture patterns and structures in the data that viewers commonly attend to, either explicitly (e.g., anomaly detection) or implicitly (e.g., reading general trends).

In this work, we compile a set of visual features that recur across multiple domains and support common analytical and perceptual tasks in line charts. We identified candidate features through a targeted review of prior work in visualization, time-series analysis, signal processing, and numerical analysis, focusing on features that are explicitly defined, repeatedly referenced, and directly interpretable from a line chart. The resulting feature set is restricted to analytically defined properties that have a clear visual representation in line charts and can be measured consistently across datasets.

Let $\{y_t\}_{t=1}^n$ denote a discrete time-series sampled at equal intervals, with index $t=1,\dots,n$. Based on this setup, we define the following visual features:

\newcommand{\visualFeatureFig}[1]{\begin{wrapfigure}[4]{L}{0.28\linewidth}
    \vspace{-8pt}
    \includegraphics[width=1.1\linewidth]{#1}
\end{wrapfigure}}

\visualFeatureFig{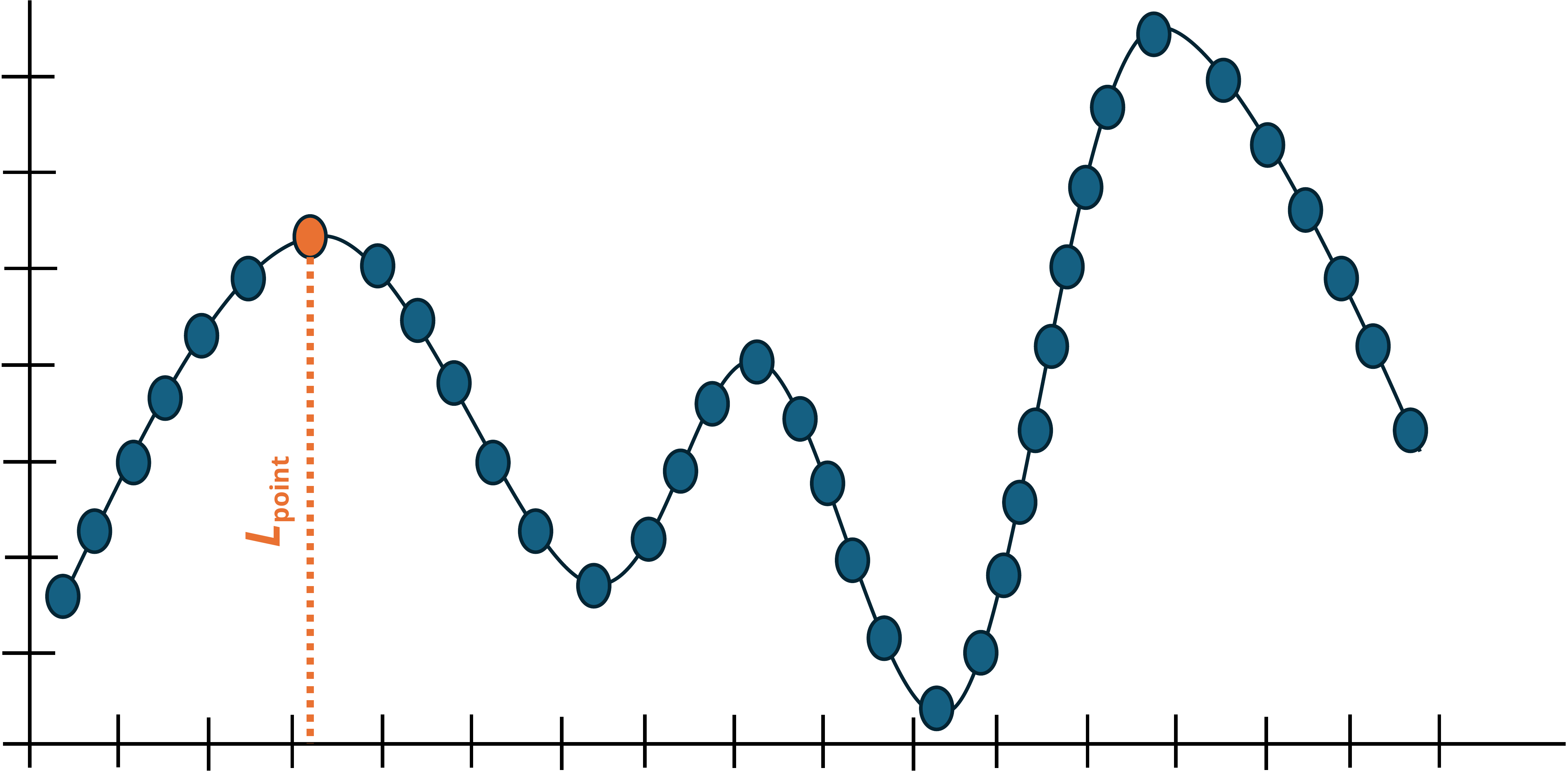}
\noindent
We define \textbf{Level} as the vertical magnitude of a series at a given moment, representing its instantaneous height or value~\cite{albers2014task}. 
This feature supports tasks such as identifying maxima or minima, where the viewer focuses on the pointwise magnitude of the data. 
Formally, for a discrete time-series $y_1, \dots, y_n$, the level at time $t_0$ is
$L_{\text{point}}(t_0) = y(t_0)$.
While prior work distinguishes an \emph{interval-level} as the average value over a span, this concept is captured globally by the \textbf{Mean} and locally by \textbf{Regimes}, and thus is not treated as a separate feature here.

\visualFeatureFig{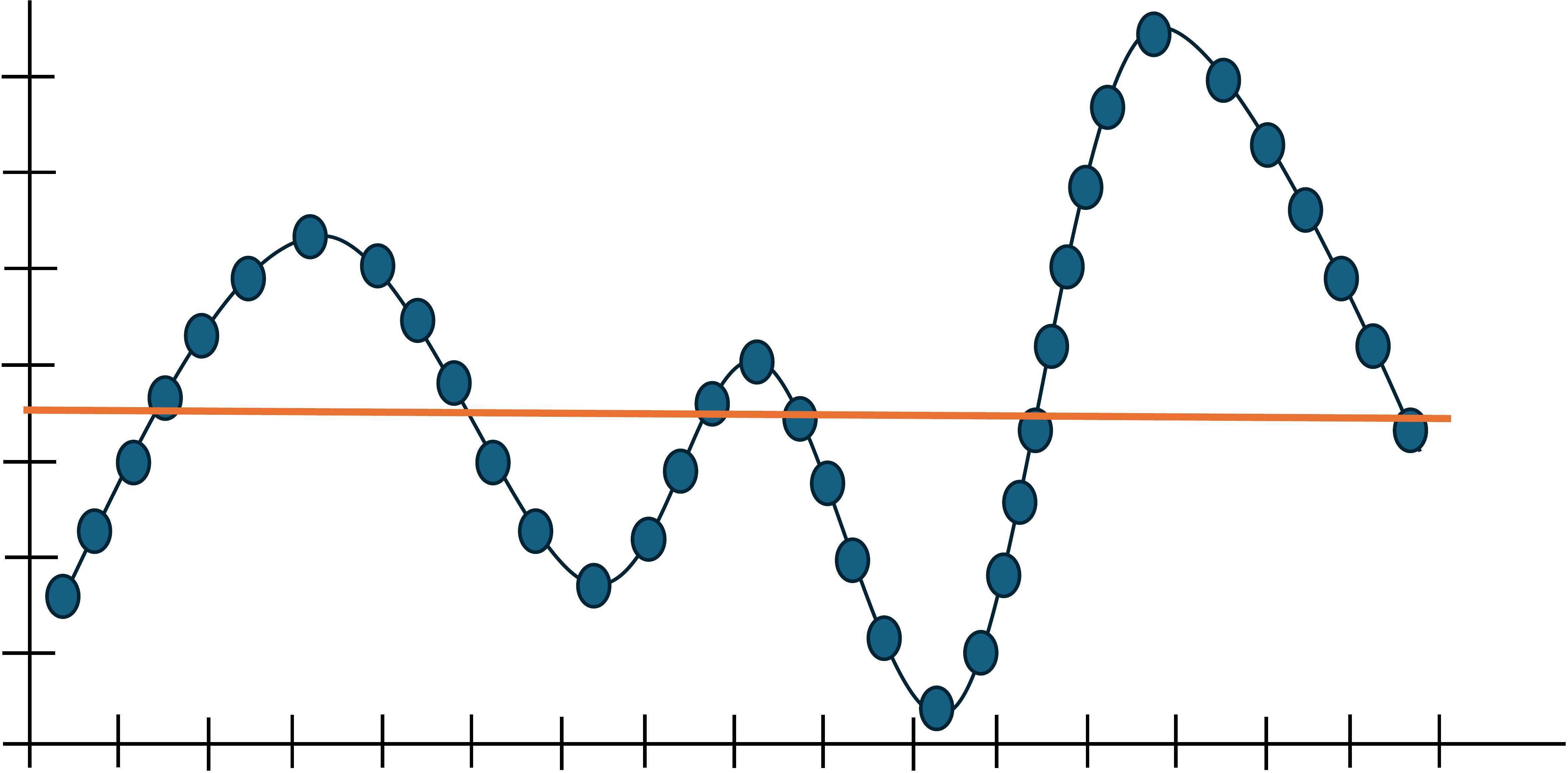}
\noindent
\textbf{Mean} represents the balance point or visual ``center'' of the data, describing the overall height around which the series fluctuates. It summarizes the central tendency of all samples.
For a discrete series $y_1,\dots,y_n$, the mean is:
$\mu \;=\; \frac{1}{n}\sum_{i=1}^{n} y_i$.

\visualFeatureFig{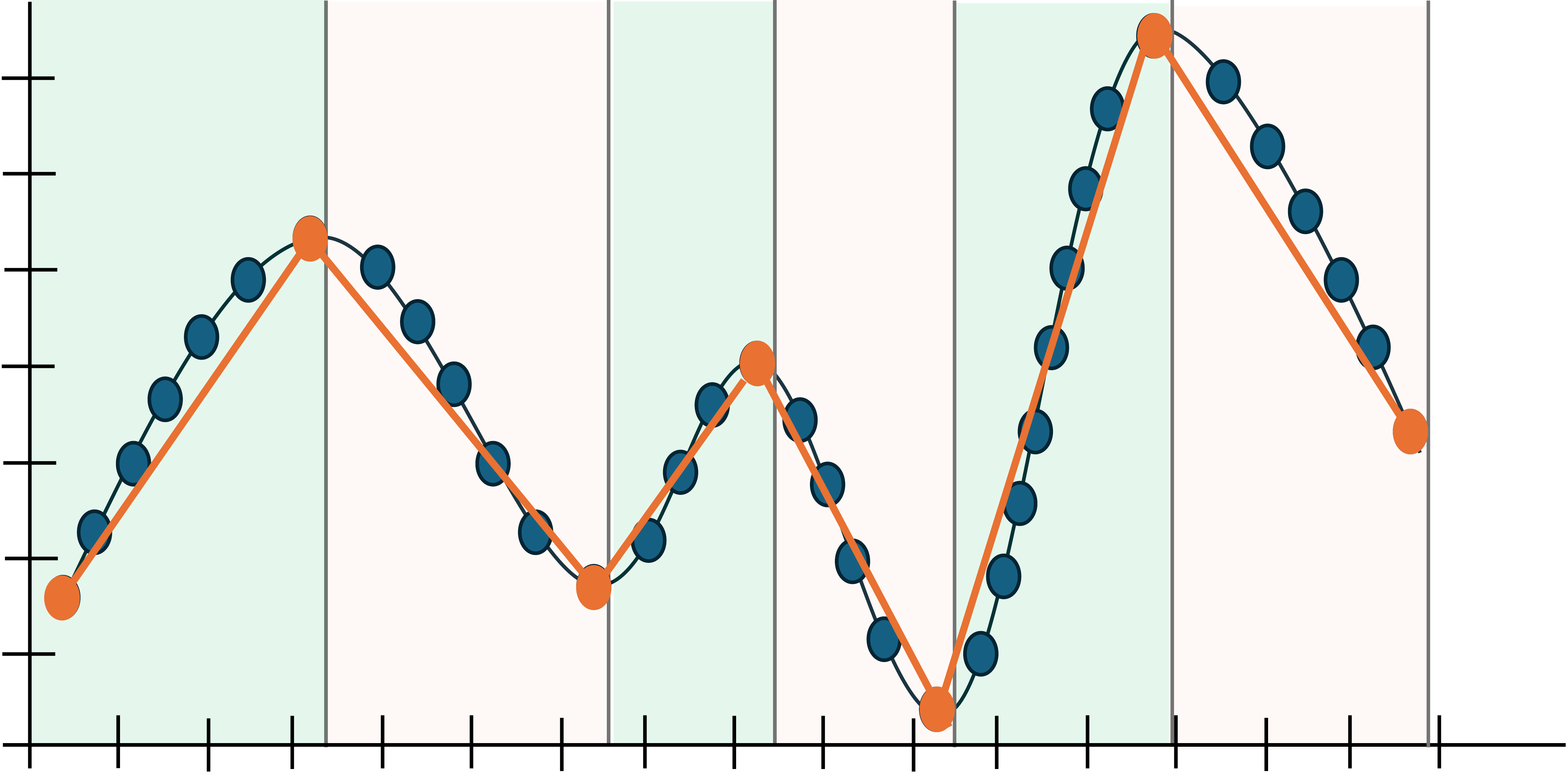}
\noindent
\textbf{Regimes} appear visually as stable, sustained phases or plateaus in which the data hovers around a consistent baseline. \textbf{Change-points} are the boundaries where these baselines shift, perceived as visible jumps or structural breaks. In time-series analysis, regimes correspond to states with distinct statistical properties~\cite{hamilton2020time}.
Formally, partition the index set into contiguous intervals $\{[a_j,b_j]\}_{j=1}^k$ with $a_1=1$, $b_k=n$, and $b_j+1=a_{j+1}$. The baseline of regime $j$ is:
\[
L_j \;=\; \frac{1}{b_j-a_j+1} \sum_{t=a_j}^{b_j} y_t.
\]
A change-point occurs at the boundary $b_j$ whenever
$L_j \;\neq\; L_{j+1}$.
Unlike local extrema (directional reversals), change-points mark shifts in underlying statistics such as mean, variance, or trend.

\visualFeatureFig{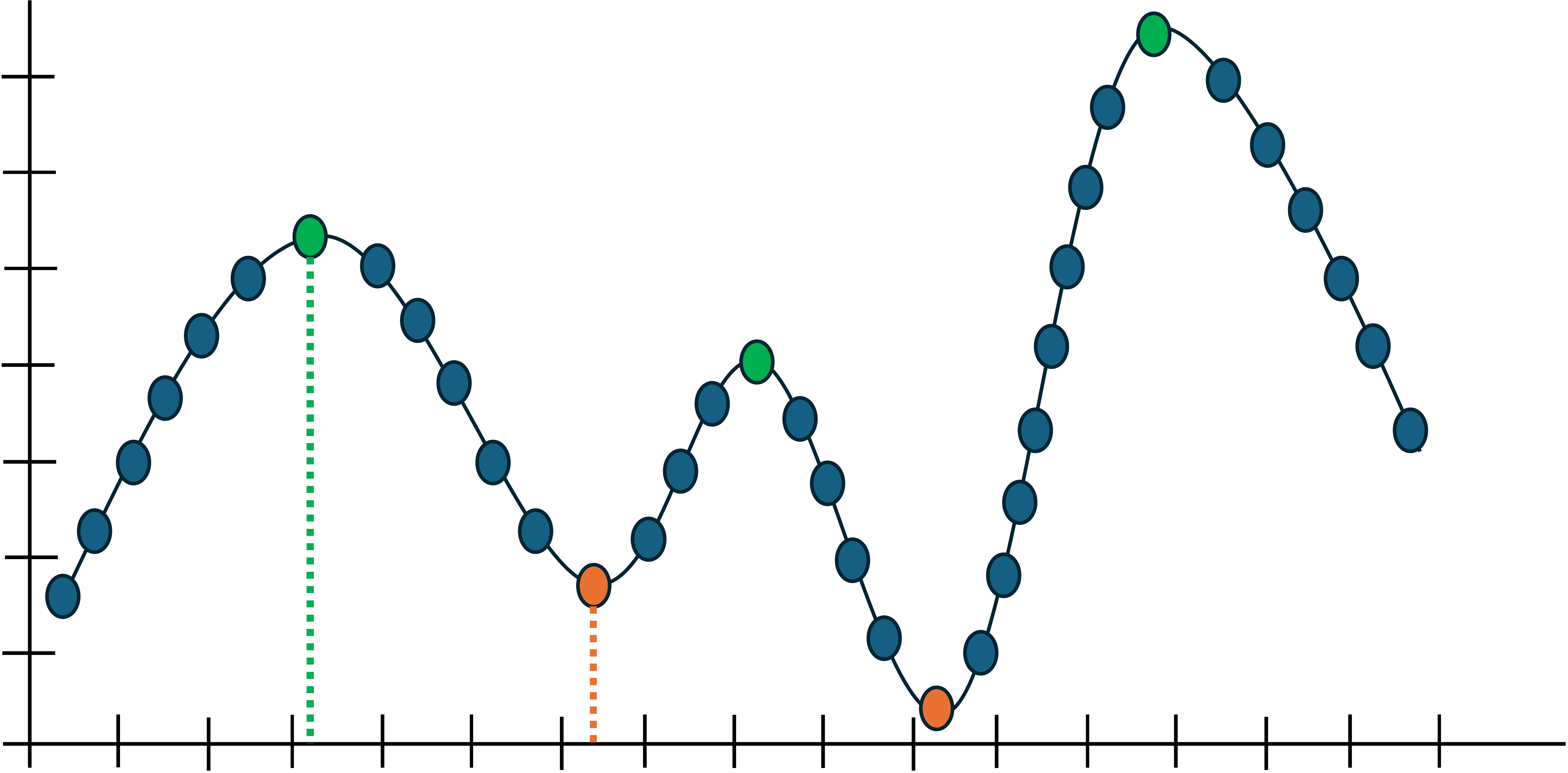}
\noindent
\textbf{Local extrema} denote visible turning points where the curve reverses direction. Extrema occur when the first derivative changes sign.
$y_t$ is a local maximum if it is greater than its immediate neighbors: 
$y_{t-1} < y_t > y_{t+1}$.
$y_t$ is a local minimum if it is smaller than its immediate neighbors:  
$y_{t-1} > y_t < y_{t+1}$.
This three-point comparison identifies every local turning point at the finest resolution the data supports.

\visualFeatureFig{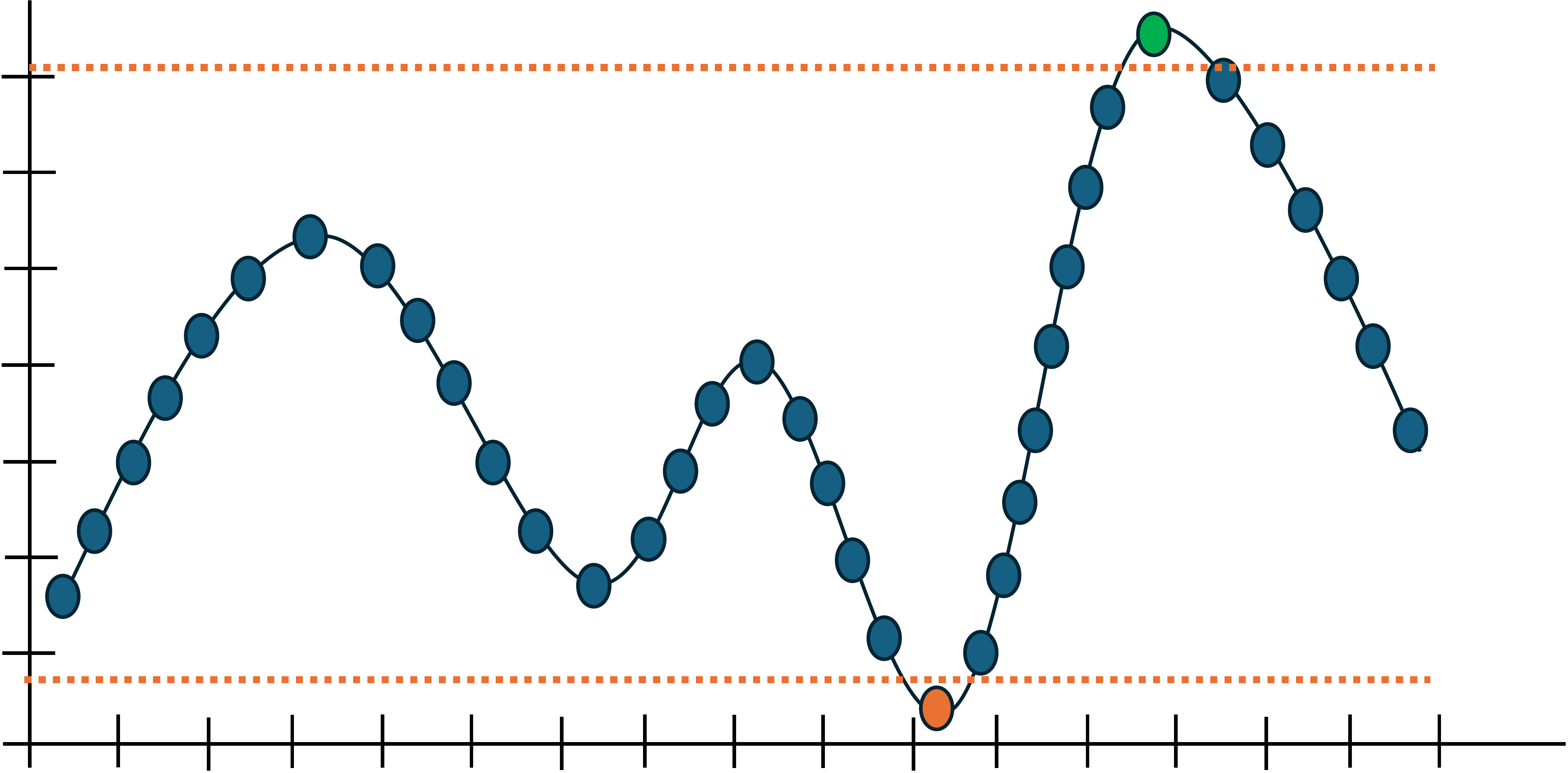}
\noindent
\textbf{Spikes} \& \textbf{Dips} denote short, high-magnitude deviations relative to the local context that often dominate visual attention~\cite{bromley2023difference}. 
We define them as local outliers detected within a sliding window. For a centered window of size $w$, let $\mu_w(t)$ denote the local mean. A spike or dip occurs when:
$|\,y_t - \mu_w(t)\,| \;>\; \tau$,
where $\tau$ is a user or domain-specified deviation threshold.

\visualFeatureFig{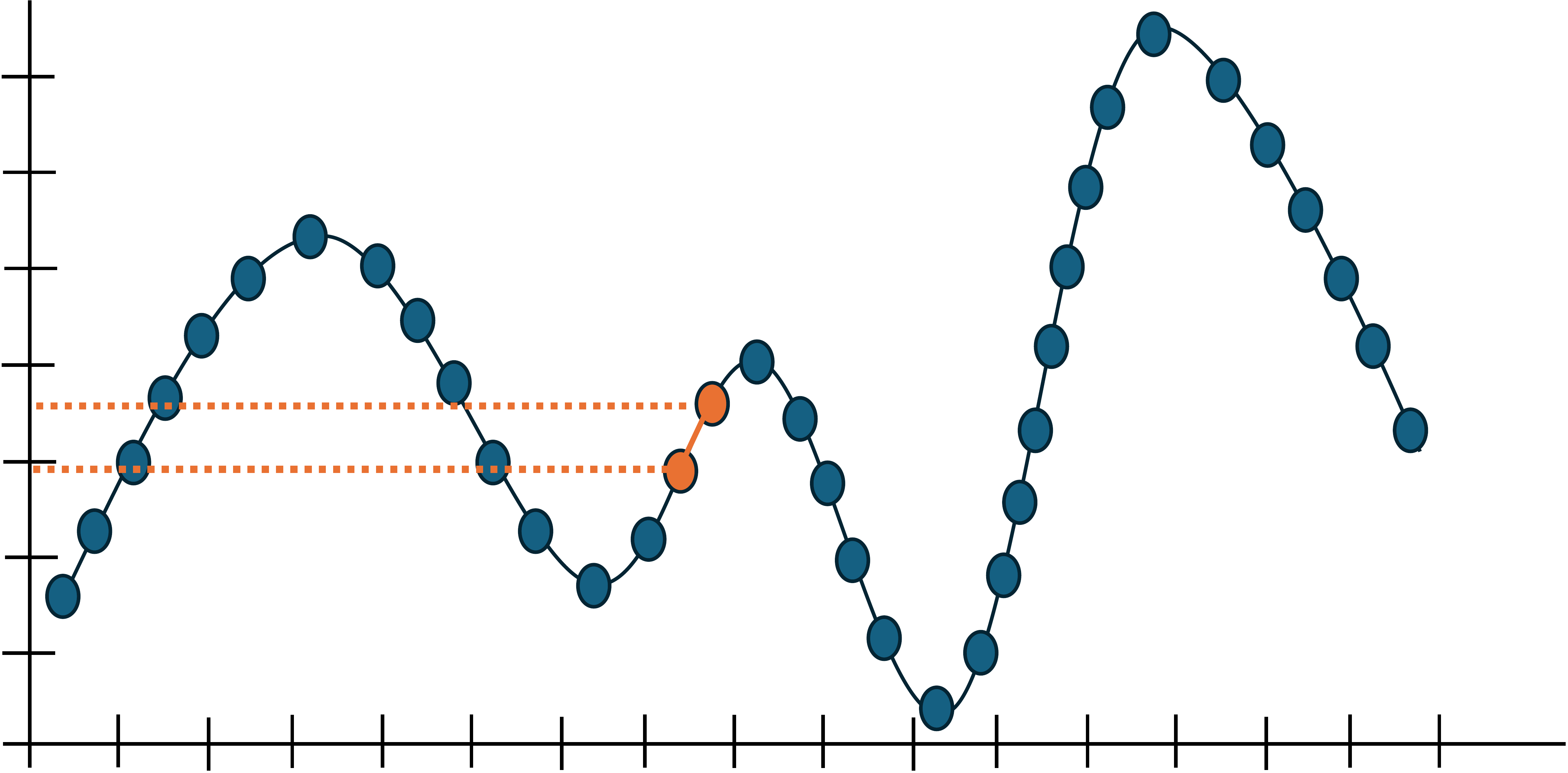}
\noindent
\textbf{Slope} conveys direction and steepness, showing how quickly the line rises or falls. It describes the local rate of change between consecutive samples.
For evenly spaced data, the slope at point \emph{t} is:
$m(t) \;=\; y_{t+1} - y_t$.

\visualFeatureFig{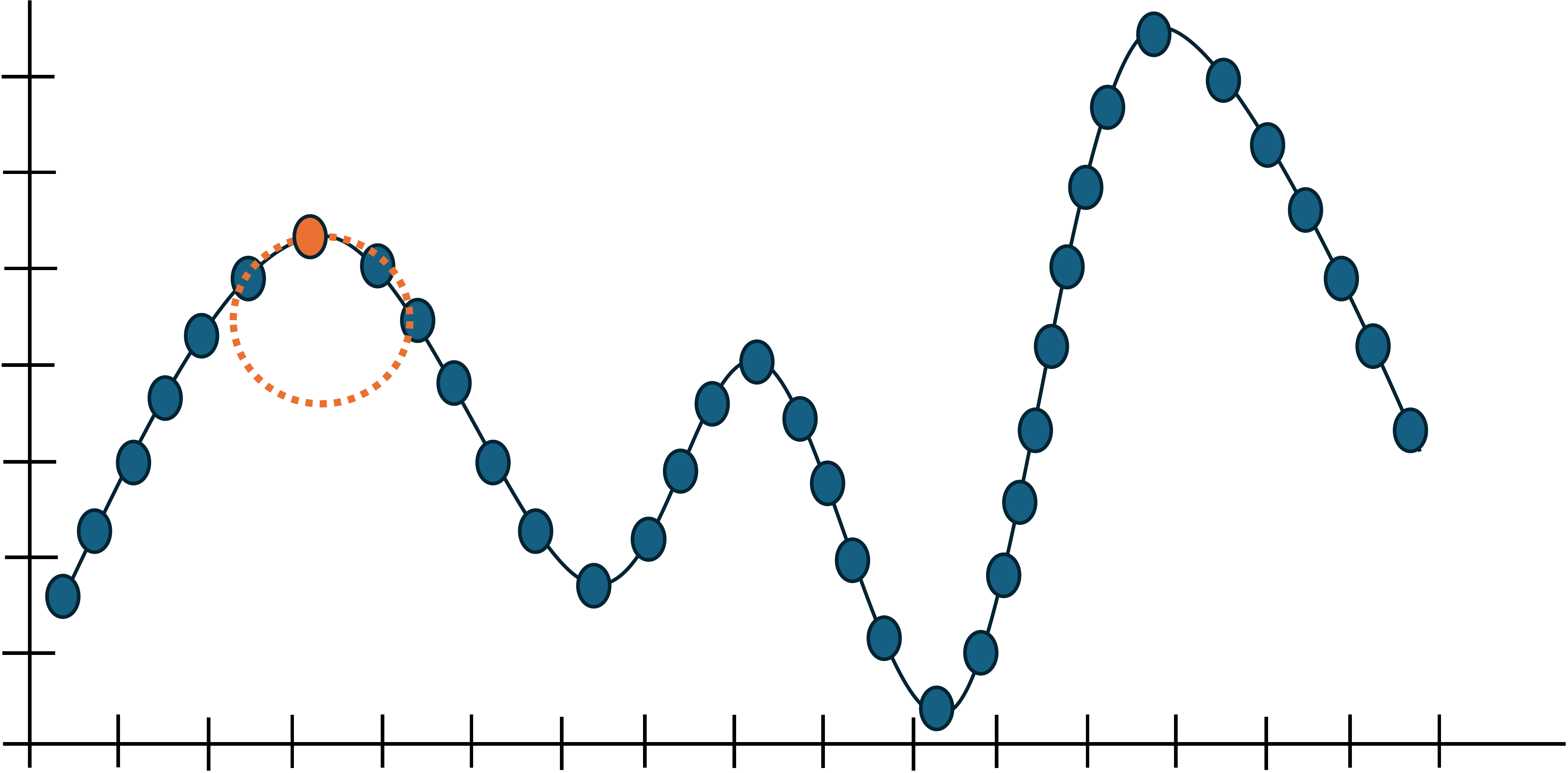}
\noindent
\textbf{Curvature} describes how sharply a curve bends at a point. It is the rate of change of the tangent direction according to differential geometry~\cite{do2016differential}. For evenly spaced data, we approximate derivatives by centered differences, yielding:
\[
\kappa_t \;\approx\;
\frac{\big|\,y_{t+1} - 2y_t + y_{t-1}\,\big|}
{\Big(1 + \big(\tfrac{y_{t+1} - y_{t-1}}{2}\big)^2\Big)^{3/2}}.
\]

\visualFeatureFig{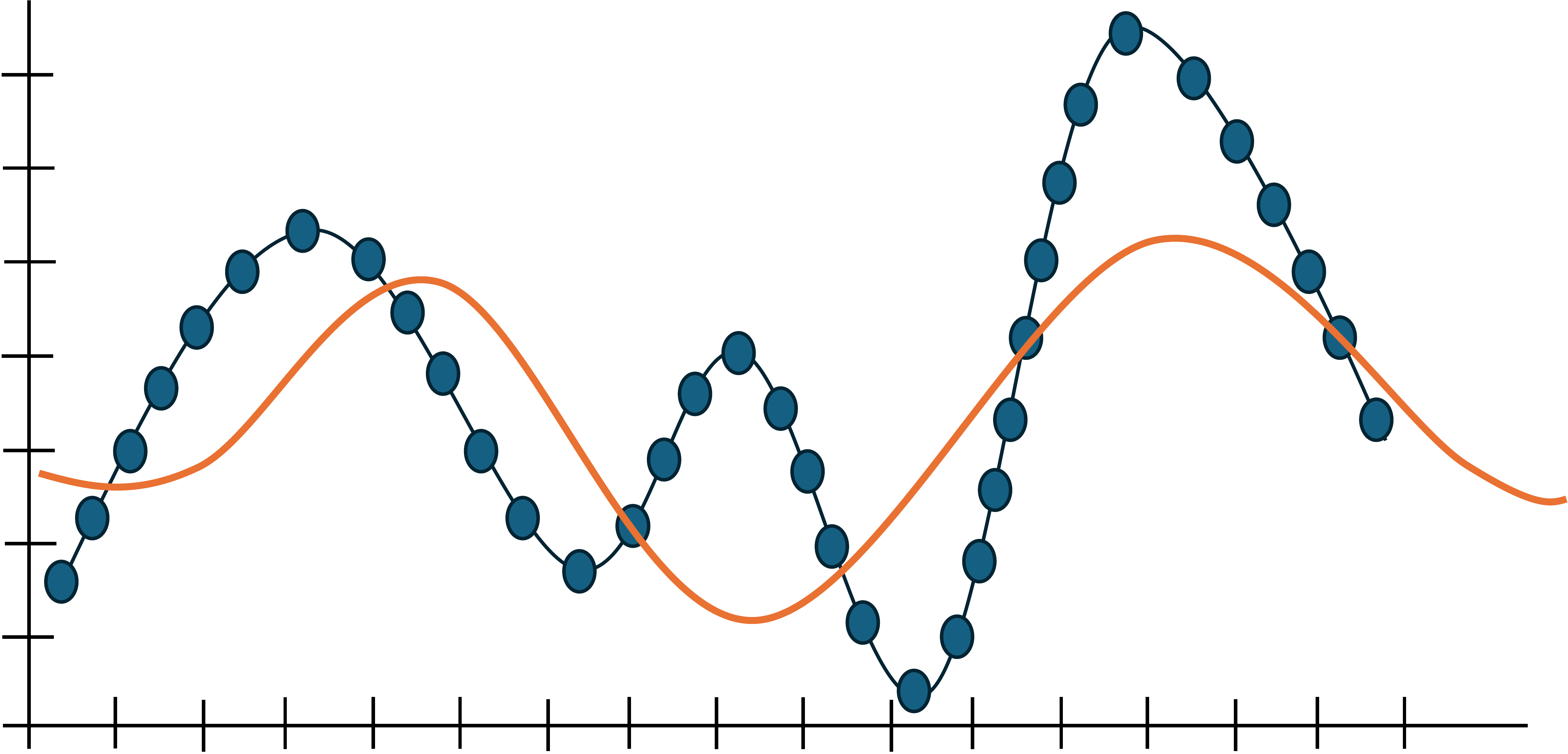}
\noindent
\textbf{Trend} captures the underlying pattern of the data. It represents the low-frequency component of a time-series, representing broad upward or downward movements over time~\cite{proma2025visual}.
Trend can be obtained by attenuating high-frequency content with a low-pass filter in the frequency domain~\cite{musbah2020novel, duda2018detection, denholm1998practical}:
$\hat{T}(t) = \text{InverseFFT}\Big( H_{\mathrm{LP}}(f)\cdot \text{FFT}(y(t)) \Big)$,
where $H_{\mathrm{LP}}(f)$ is a low-pass filter that keeps frequencies $|f|\le f_c$ and removes higher ones.

\visualFeatureFig{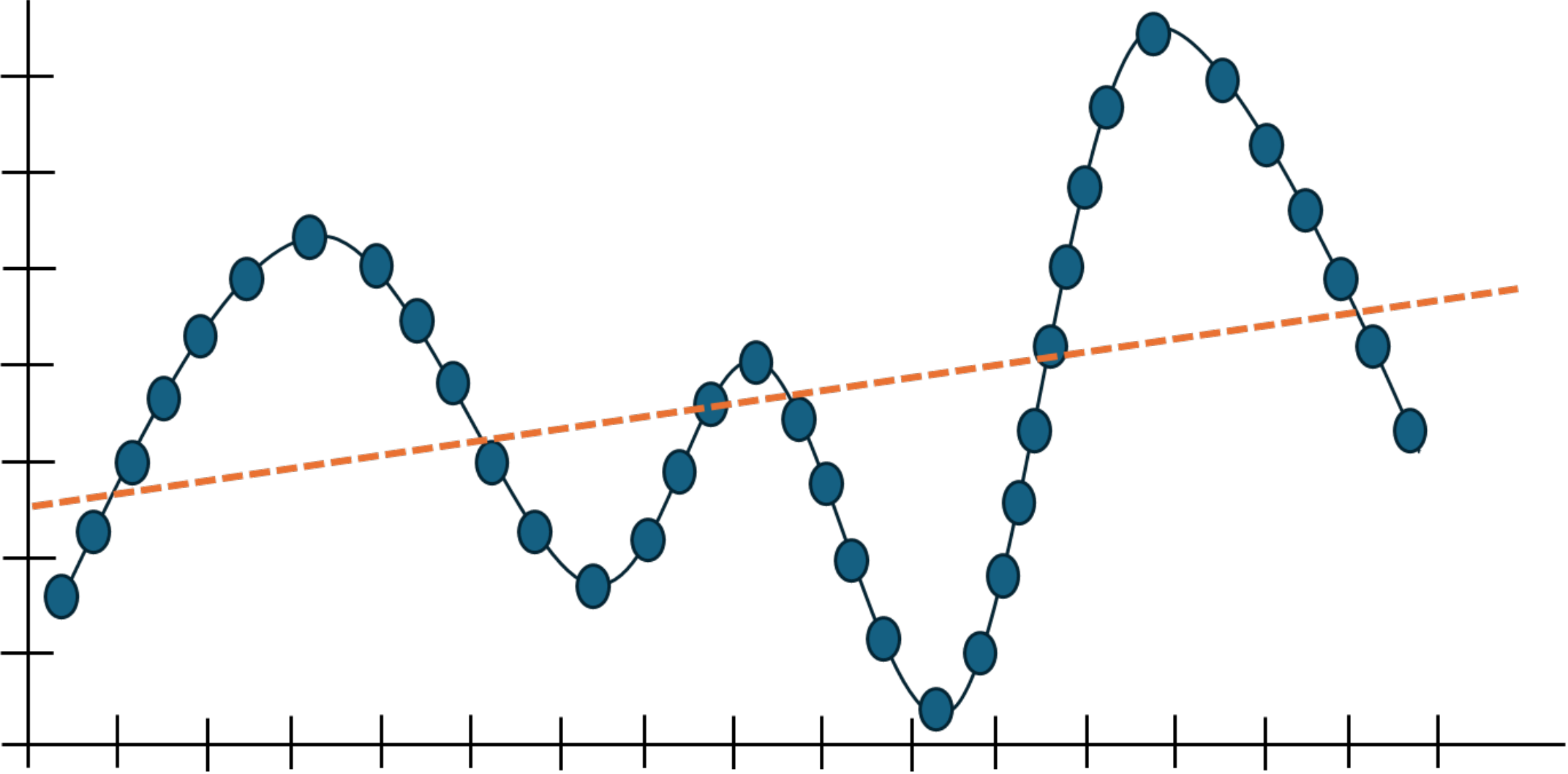}
\noindent
\textbf{Regression fit} provides a simplified, model-based depiction of overall direction and rate of change. It expresses the average trend through a parametric form, commonly an ordinary least-squares (OLS) line~\cite{correll2017regression, cleveland1979robust}.
The OLS regression fit line can be defined as:
${y}_t = {\alpha} + {\beta}t$,
where the slope ${\beta}$ gives the average rate of increase or decrease and the intercept ${\alpha}$ anchors the line vertically.

\visualFeatureFig{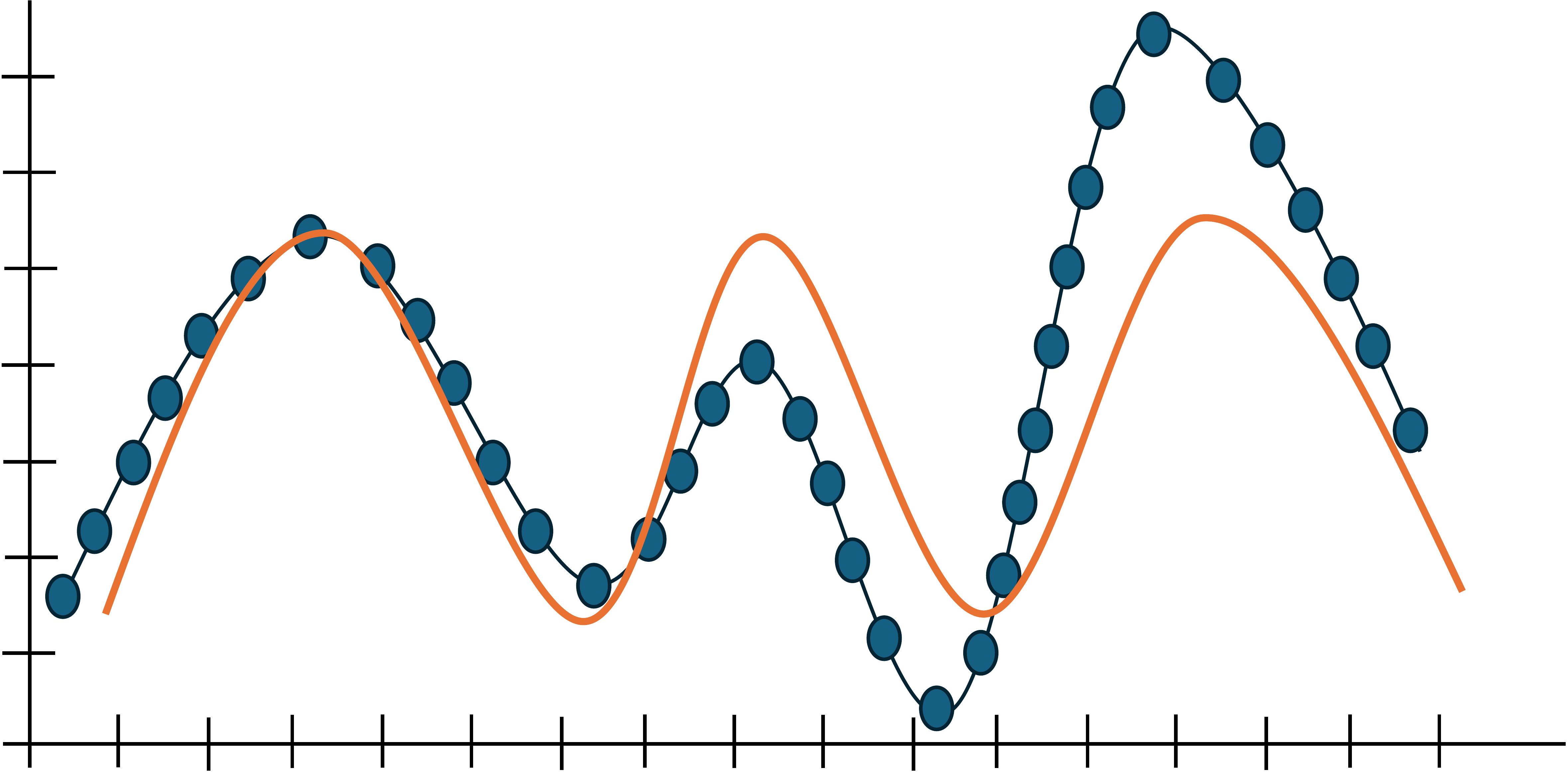}
\noindent
\textbf{Periodicity} refers to repeating cycles or oscillations that recur at a regular rate. Prior studies have detected periodicity with Fast Fourier Transformation~\cite{proma2025visual,musbah2020novel,wen2023robust}. 
In the frequency domain,
$P(f) = |\text{FFT}(y(t))|$.
If $f^*$ is the dominant nonzero frequency, the corresponding period and amplitude are
$T = \frac{1}{f^*}, \qquad A = P(f^*)$.

\visualFeatureFig{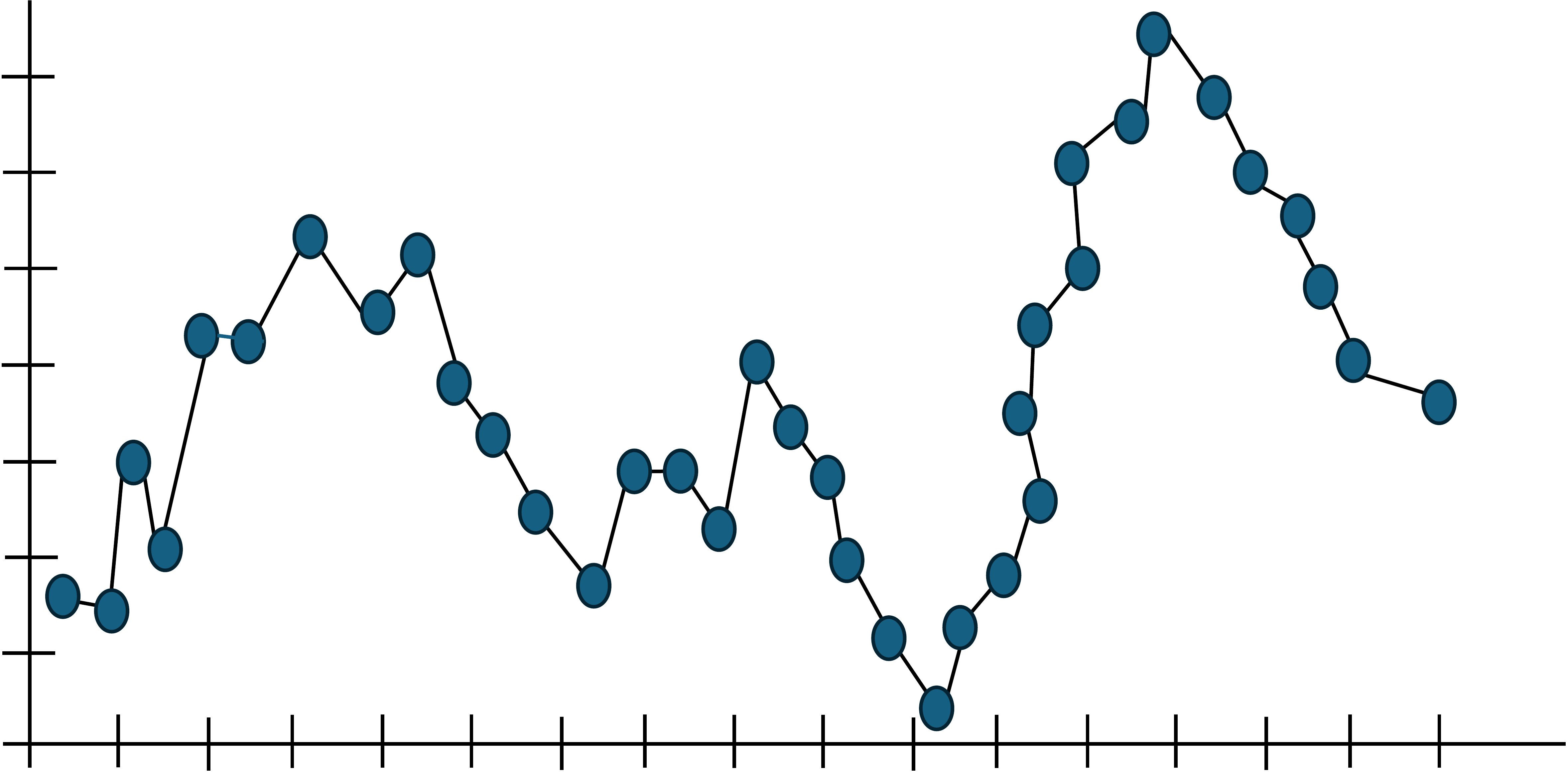}
\noindent
\textbf{Roughness} measures how jagged or irregular the line appears. It is the small-scale ups and downs that give it a textured look. Following Rong and Bailis~\cite{rong2017asap}, we define roughness as the standard deviation of first differences in a uniformly sampled time-series, capturing the degree of local jaggedness. 
For values $Y=\{y_1,\dots,y_N\}$,
$\Delta Y = \{\,y_{i+1}-y_i : i=1,\dots,N-1\}$,
and the roughness is
$\text{roughness}(Y) = \sigma(\Delta Y)$.

\visualFeatureFig{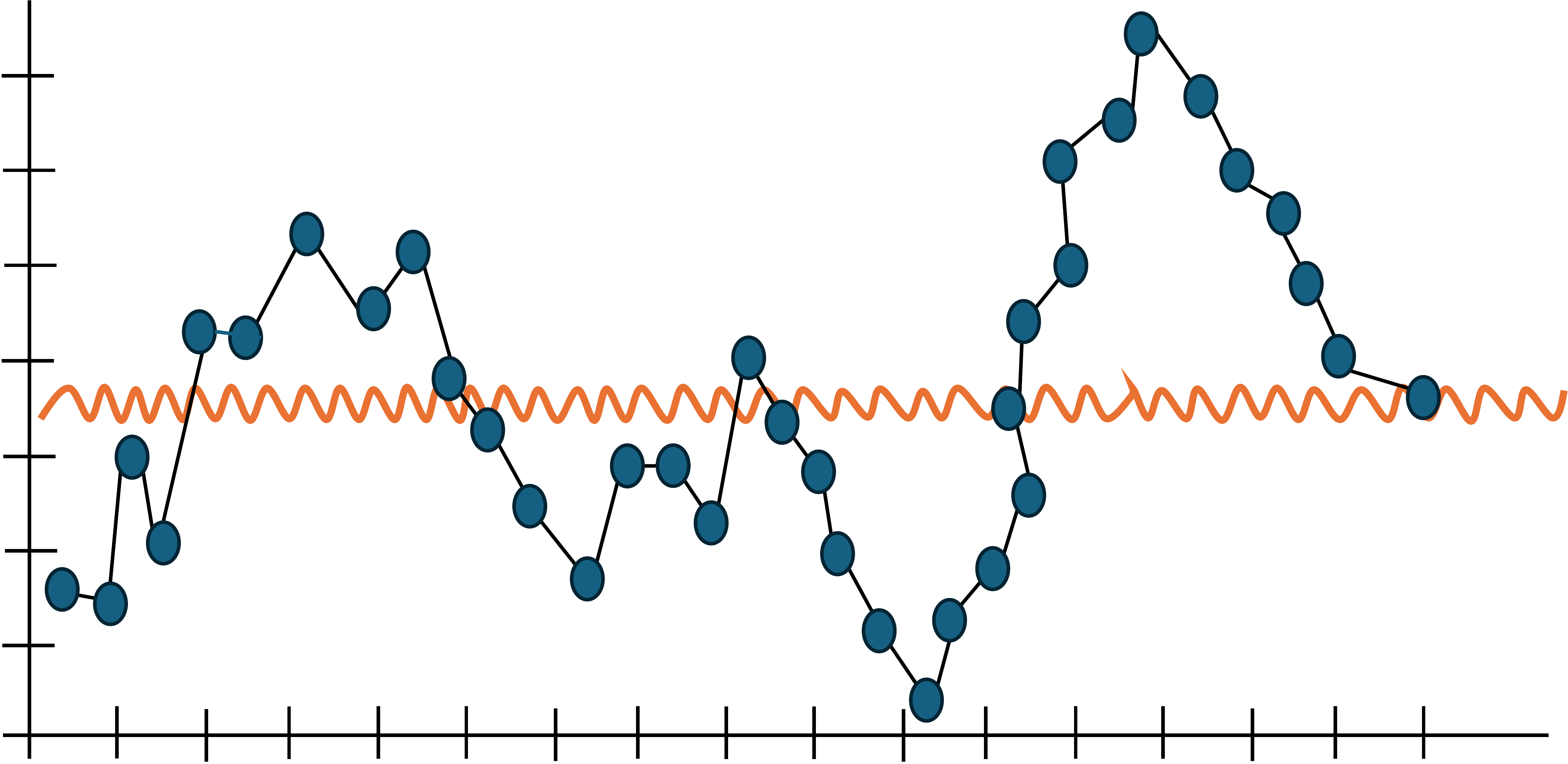}
\noindent
\textbf{Noise} represents random, high-frequency fluctuations that obscure underlying structure~\cite{proma2025visual}. In the frequency domain, noise corresponds to components concentrated at high frequencies, which can be isolated using a high-pass filter~\cite{butterworth1930theory,rhodes1980generalized,musbah2019identifying}. Formally, we decompose a series into a smooth signal or trend $T(t)$ and noise $N(t)$:
$y(t) = T(t) + N(t)$,
with the noise estimated as
$\hat{N}(t) = \text{InverseFFT}\Big( H_{\mathrm{HP}}(f) \cdot \text{FFT}(y(t)) \Big)$,
where $H_{\mathrm{HP}}(f)$ retains frequencies $|f| > f_c$.

%% file: sec.experimental_setup.tex
\section{Experimental Setup}
\label{sec:experimental_setup}

We evaluate line chart simplification techniques by applying each method to a fixed collection of real-world time-series over a range of simplification degrees. For every technique-dataset pair, we generate a large set of simplified outputs that serve as the basis for feature preservation analysis.

\subsection{Algorithm Implementation}
We implement all the algorithms using standard formulations from the original
papers~\cite{jugel2014m4,steinarsson2013downsampling,van2023minmaxlttb,li2024fpcs,rong2017asap}.
Many of these algorithms depend on multiple parameters. For example, \methodSG depends on window size and polynomial degree, while \methodBW, \methodCheb, and \methodEllip take parameters such as cutoff frequency, filter order, and ripple. Varying all such parameters jointly would create a high-dimensional space that is difficult to interpret and makes it impossible to align
simplification degrees across algorithms. To ensure comparability and a monotonic simplification degree per algorithm, we vary only one primary parameter per method: the parameter that most directly governs simplification degree, such as window size for sliding-window filters, cutoff frequency for IIR filters, output length for fixed-count reducers, rate for \methodFPCS{}, bin size for \methodPAA and resolution for \methodASAP{}, while keeping all other parameters fixed at reasonable default values (e.g., filter order~$=2$, polyorder~$=2$, ripple~$=0.5$). This design follows LineSmooth's~\cite{rosen2020linesmooth} approach of sweeping a single smoothing parameter while holding secondary parameters constant. 
Time-complexity per algorithm is reported in~\Cref{tab:data_simplification_categories}.

\subsection{Sampling Across Simplification Degrees}
For each algorithm, we vary its main simplification parameter across 100 values to span a range from minimal to heavy simplification. This sampling is performed over the algorithm’s parameter values, while keeping the input data fixed. This sampling strategy provides a sufficiently fine resolution for observing how feature preservation is affected across different simplification degrees. To produce different simplification degrees, we sample parameters on either a linear or logarithmic scale, depending on how sensitively the algorithm responds to parameter changes. Parameters whose effect on simplification grows non-linearly (e.g., cutoff frequencies of IIR and FFT filters, window sizes of sliding-window filters, or desired output length of reducer algorithms) are sampled on a logarithmic scale, while parameters with an approximately linear effect (e.g., $\sigma$ in \methodGau filter and rate in \methodFPCS{}) are sampled linearly. This strategy ensures that the sampled parameter values progress systematically.

\begin{table}[!b]
\centering
\caption{Summary of dataset domains, dataset groups, and structural variation 
(80 datasets from LineSmooth~\cite{rosen2020linesmooth}).}
\label{tab:datasets_overview}
\resizebox{\linewidth}{!}{
\begin{tabular}{p{0.22\linewidth} p{0.45\linewidth} p{0.55\linewidth} p{0.10\linewidth}}
\toprule
\textbf{Domain} & \textbf{Dataset Groups} & \textbf{Description} & \textbf{Length} \\
\midrule

Astronomy &
\begin{tabular}[t]{@{}l@{}}
\texttt{astro}
\end{tabular} &
5 data files; with frequency measurement of radio waves emitted from extraterrestrial matter~\cite{alma} &
1947 \\
\midrule

Climate &
\begin{tabular}[t]{@{}l@{}}
\texttt{climate\_awnd}\\
\texttt{climate\_tmax}\\
\texttt{climate\_prcp}
\end{tabular} &
18 data files; includes average wind speed (awnd), high temperature (tmax), and total precipitation (prcp) from 6 U.S. airports~\cite{noaa} &
3651 \\
\midrule

Crime &
\begin{tabular}[t]{@{}l@{}}
\texttt{chi\_homicide\_weekly}\\
\texttt{chi\_homicide\_monthly}
\end{tabular} &
2 data files; weekly and monthly homicide counts in Chicago from January 2001 through July 2019~\cite{chicago_crime} &
969  222 \\
\midrule

EEG &
\begin{tabular}[t]{@{}l@{}}
\texttt{eeg\_500}\\
\texttt{eeg\_2500}\\
\texttt{eeg\_10000}
\end{tabular} &
18 data files; EEG signals from 6 channels per window length (500, 2500, 10000) recorded during a visual attention task~\cite{eeg_database} &
500  2500  10000 \\
\midrule

Flights &
\begin{tabular}[t]{@{}l@{}}
\texttt{flights\_daily}\\
\texttt{flights\_weekly}\\
\texttt{flights\_monthly}
\end{tabular} &
3 data files; daily, weekly, and monthly U.S. flight volumes from January 1, 1988 through December 31~\cite{d3_zoomable} &
7671  1095  252 \\
\midrule

New Zealand Tourism &
\begin{tabular}[t]{@{}l@{}}
\texttt{nz\_tourist\_monthly}\\
\texttt{nz\_tourist\_annual}
\end{tabular} &
2 data files; monthly and annual tourist counts from April 1921 through April 2018~\cite{new_zealand} &
1165  96 \\
\midrule

Stock Price &
\begin{tabular}[t]{@{}l@{}}
\texttt{stock\_price}
\end{tabular} &
9 data files; daily closing prices for major company stocks~\cite{yahoo_finance} &
1257 \\
\midrule

Stock Volume &
\begin{tabular}[t]{@{}l@{}}
\texttt{stock\_volume}
\end{tabular} &
9 data files; daily trading volumes for the same company stocks\cite{yahoo_finance} &
1257 \\
\midrule

Unemployment &
\begin{tabular}[t]{@{}l@{}}
\texttt{unemployment}
\end{tabular} &
14 data files; monthly unemployment counts across U.S. economic sectors from January 2000 through February 2010~\cite{us_bls} &
122 \\
\bottomrule
\end{tabular}}
\end{table}

\subsection{Datasets}
We evaluate all the algorithms for feature preservation on the same 80 real-world time-series datasets used in LineSmooth~\cite{rosen2020linesmooth}, spanning nine application domains. These datasets exhibit a wide variety of temporal behaviors such as long-term trends, strong seasonal cycles, regime shifts, noise-dominated regions, and spikes \& dips, which correspond to the visual features examined in our study. This diversity makes the collection well-suited for evaluating feature preservation under simplification.~\Cref{tab:datasets_overview} summarizes the dataset domains, their high-level groups, the number of files in each group, and their sequence lengths.

\begin{figure}[!t]
    \captionsetup[subfloat]{labelfont=scriptsize,textfont=scriptsize}
    \centering

    \subfloat[Radio astronomy spectral line]{
        \includegraphics[width=0.28\linewidth]{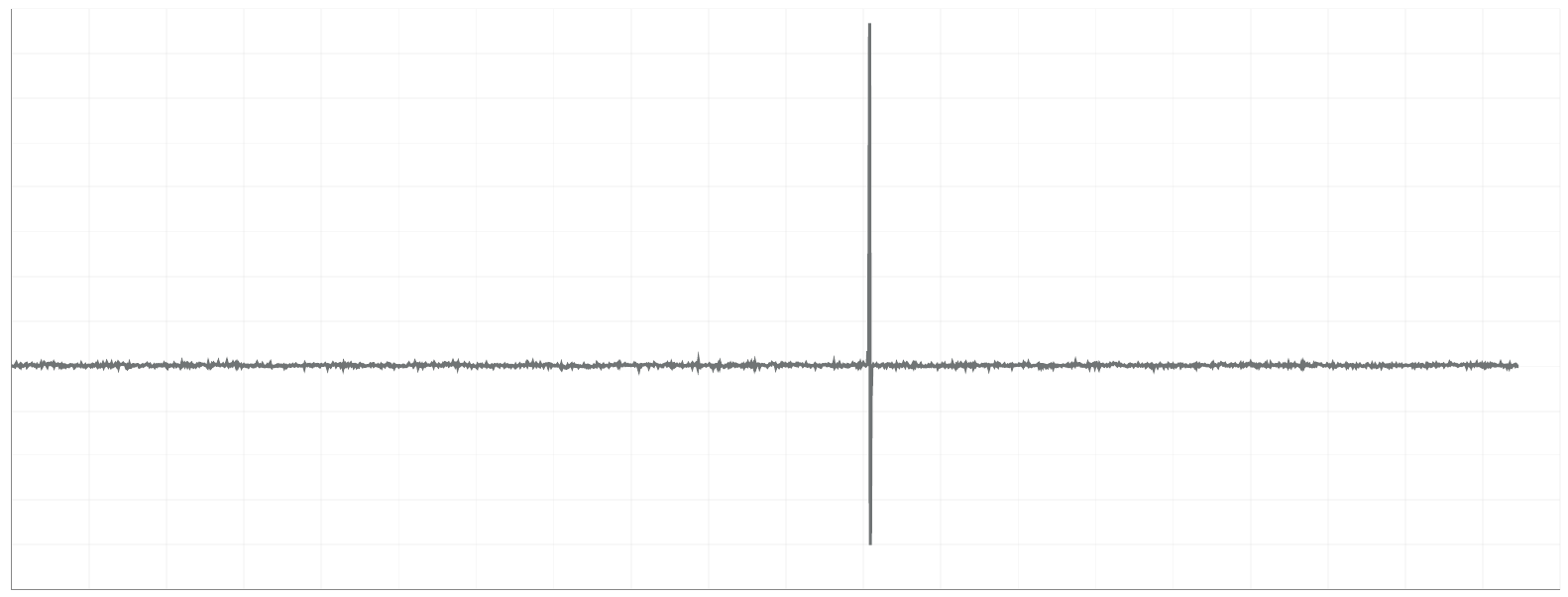}
    }
    \hspace{5pt}
    \subfloat[Daily high temperature at ORD]{
        \includegraphics[width=0.28\linewidth]{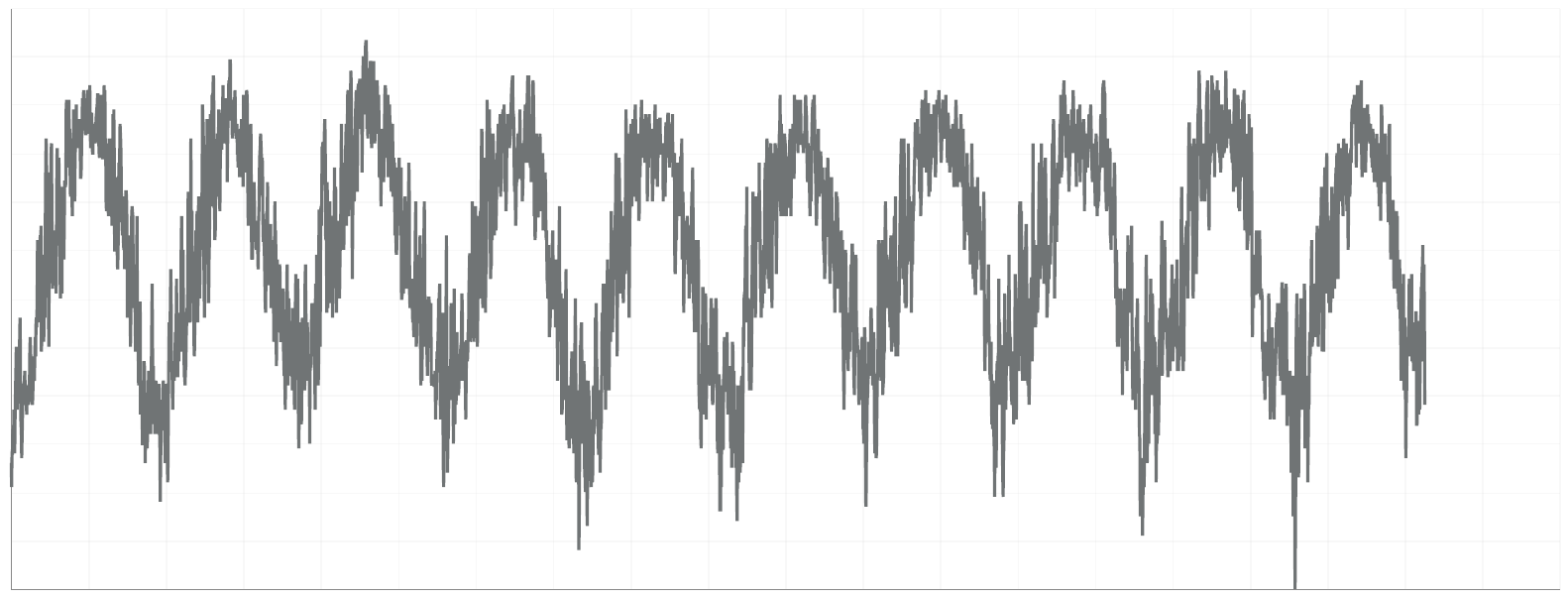}
    }
    \hspace{5pt}
    \subfloat[Weekly homicide counts in Chicago]{
        \includegraphics[width=0.28\linewidth]{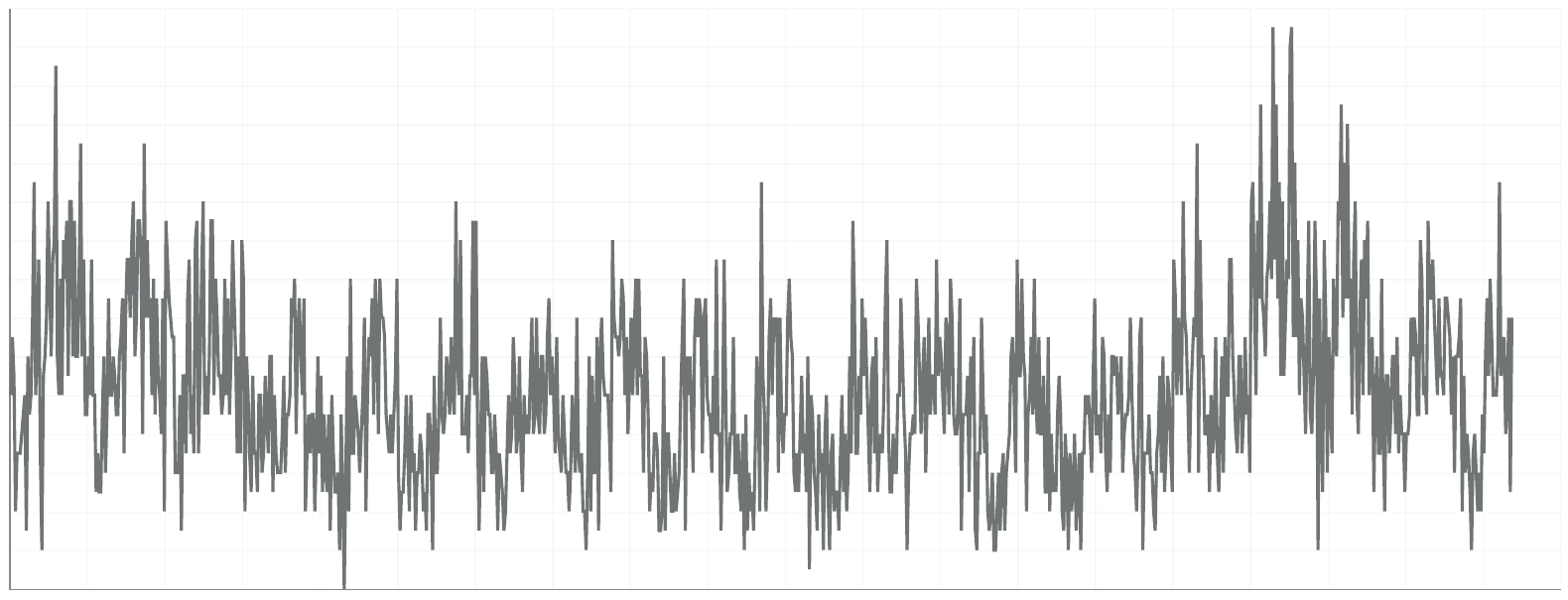}
    }

    \subfloat[EEG signal (channel 30, 10000 samples)]{
        \includegraphics[width=0.28\linewidth]{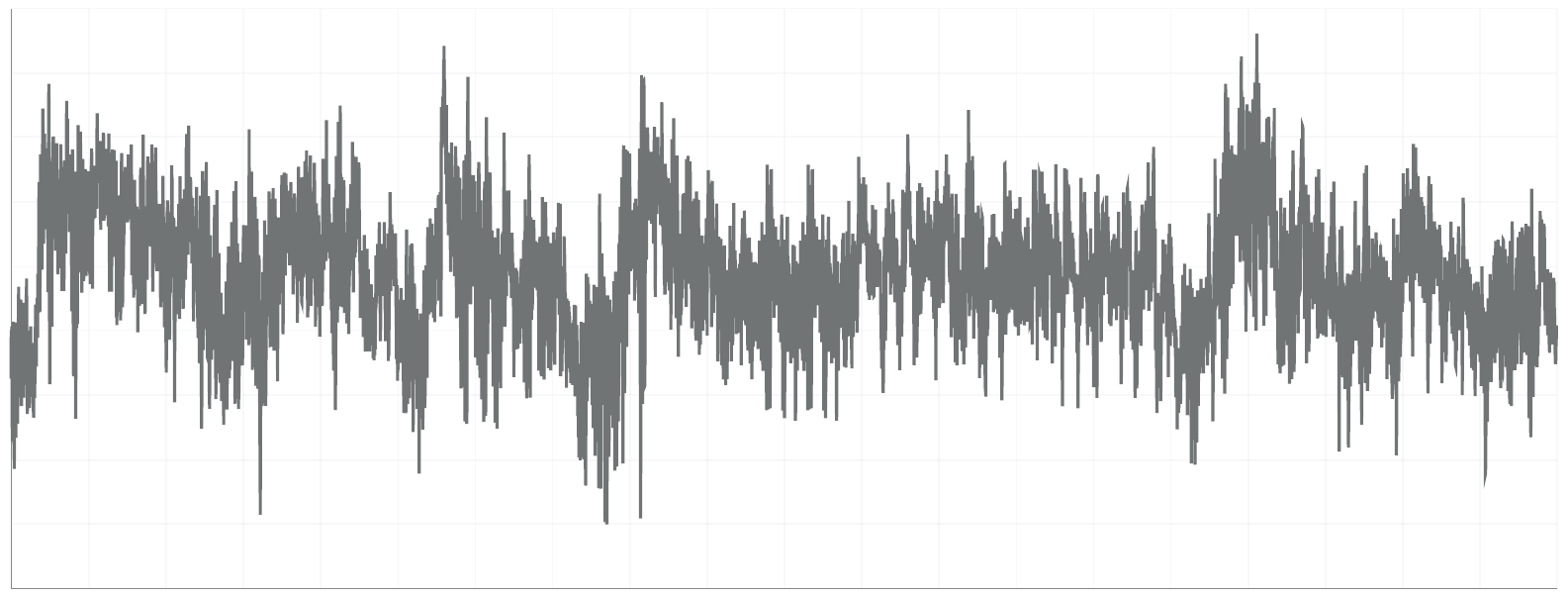}
    }
    \hspace{5pt}
    \subfloat[Weekly U.S. domestic flights]{
        \includegraphics[width=0.28\linewidth]{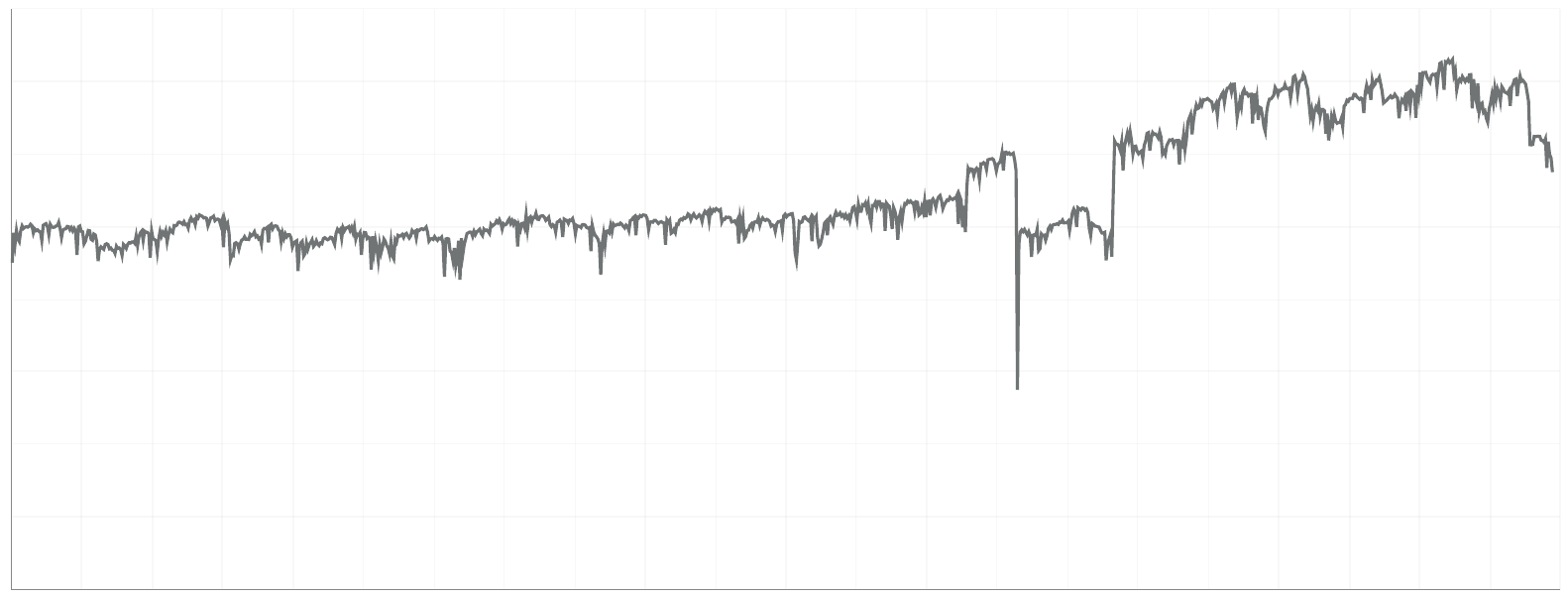}
    }
    \hspace{5pt}
    \subfloat[Monthly New Zealand tourist arrivals]{
        \includegraphics[width=0.28\linewidth]{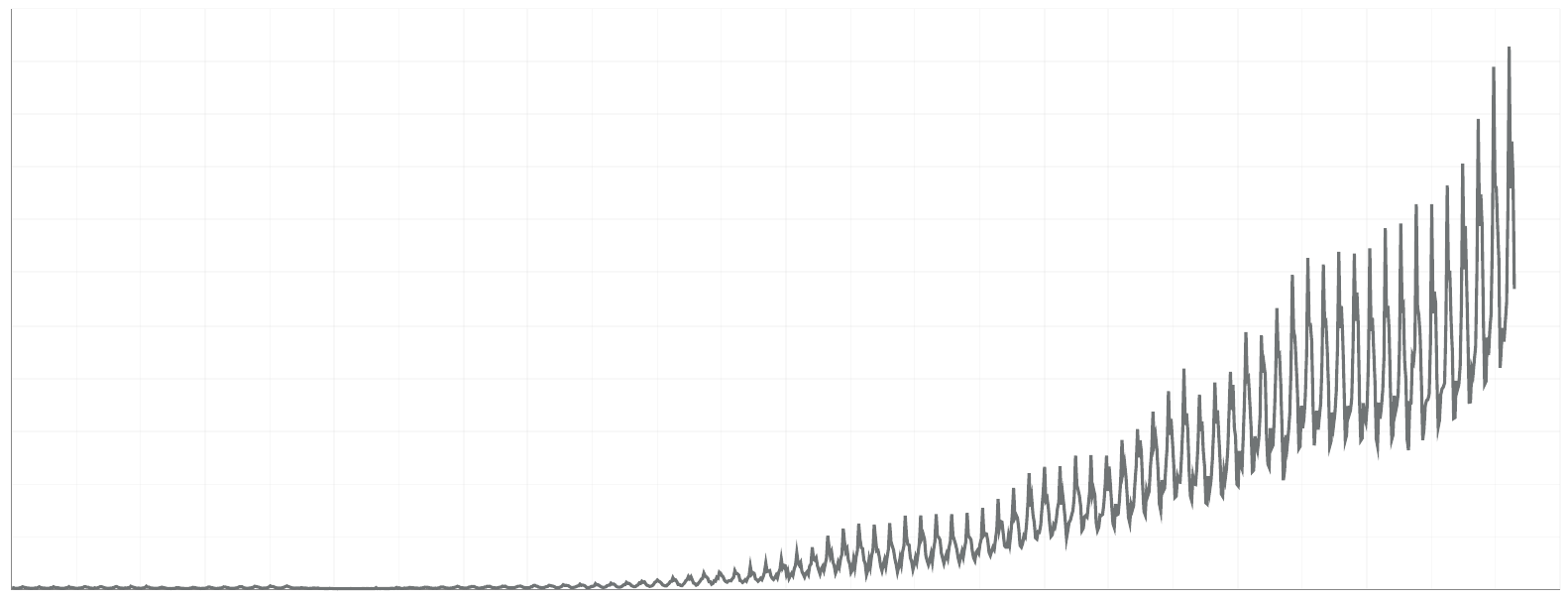}
    }

    \subfloat[Daily Apple stock closing price]{
        \includegraphics[width=0.28\linewidth]{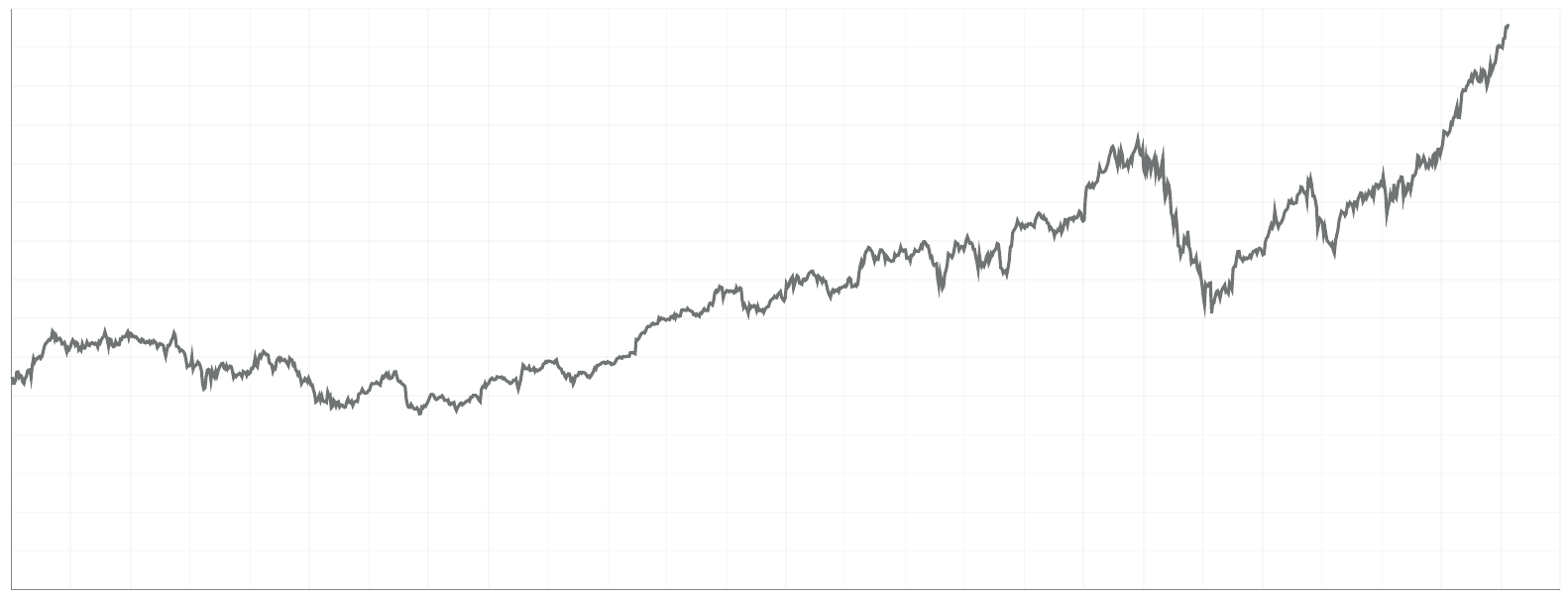}
    }
    \hspace{5pt}
    \subfloat[Daily Apple trading volume]{
        \includegraphics[width=0.28\linewidth]{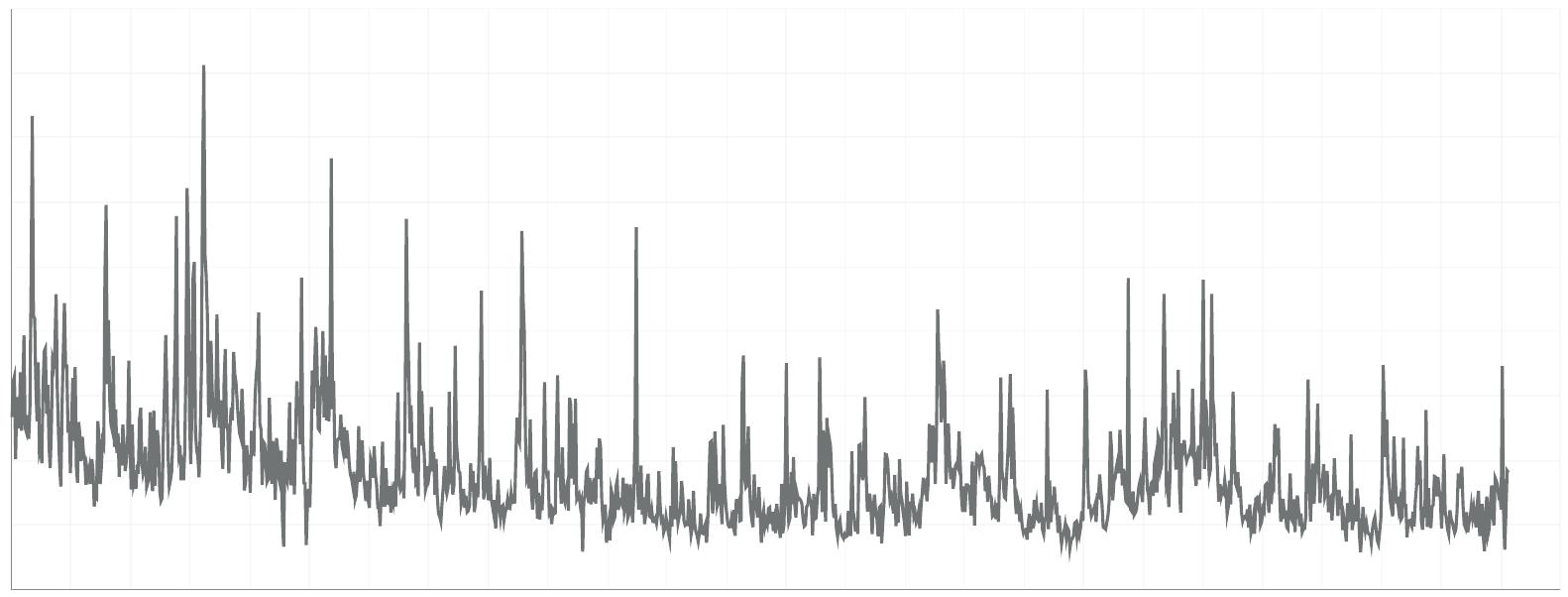}
    }
    \hspace{5pt}
    \subfloat[Monthly unemployment in hospitality sector]{
        \includegraphics[width=0.28\linewidth]{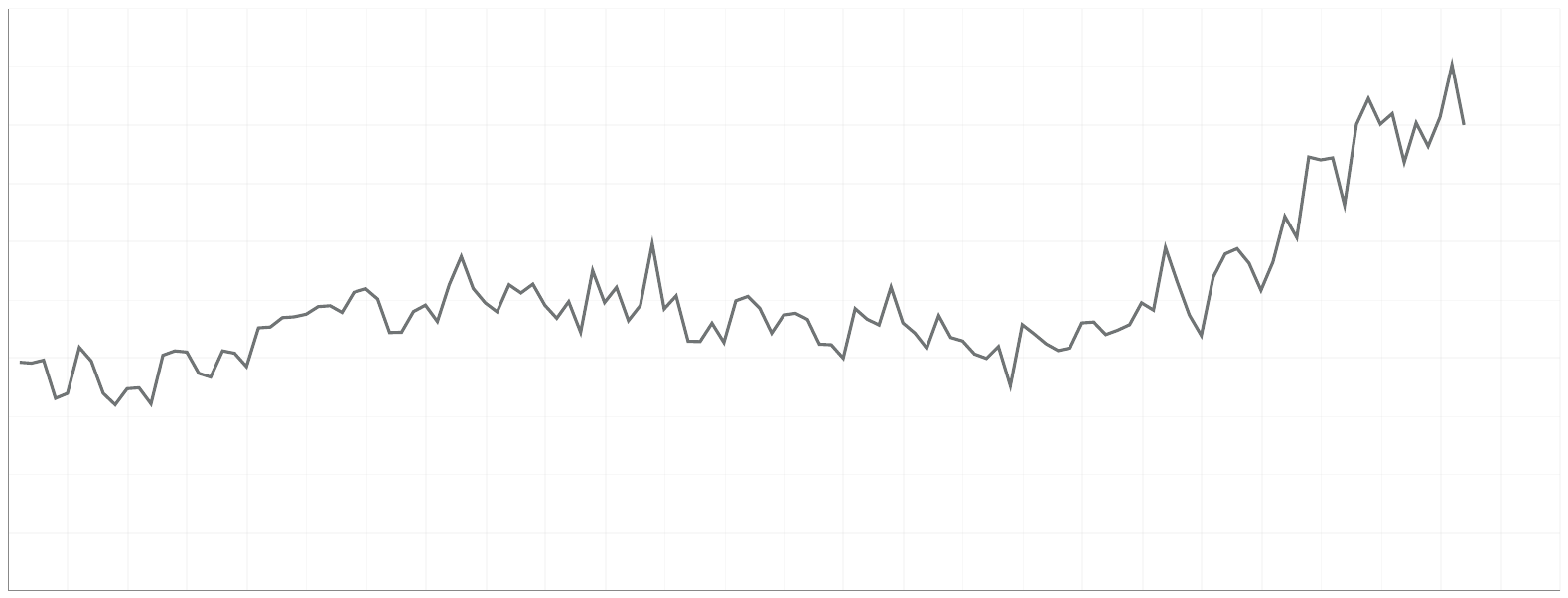}
    }

    \caption{Examples of input datasets used in our evaluation, spanning nine application domains.}
    \label{fig:original_dataset}
\end{figure}

To illustrate the diversity of structural patterns present in these datasets, \Cref{fig:original_dataset} shows one representative dataset from each domain. These examples highlight the range of visual features defined in our study: noise-dominated neural signals (\texttt{eeg\_chan30\_10000}, 10000 points), strong periodicity in climate measurements (\texttt{climate\_ord\_tmax}, 3651 points), long-term trends with local fluctuations in stock prices (\texttt{stock\_aapl\_price}, 1257 points), seasonal trends in employment data (\texttt{unemployment\_hospitality}, 122 points), combination of dip and periodic patterns in aviation data (\texttt{flights\_weekly}, 1095 points), periodicity with an upward direction in tourism (\texttt{nz\_tourist\_monthly}, 1165 points), irregular spikes \& dips in crime data (\texttt{chi\_homicide\_weekly}, 969 points), prominent extrema structures in astronomical spectra (\texttt{astro\_115\_120}, 1947 points), and heavy-tailed fluctuations in trading volume (\texttt{stock\_aapl\_volume}, 1257 points). Together, these datasets span a wide variety of visual features across domains, providing a strong basis for evaluating how simplification affects visual feature preservation. Visualizations for all 80 datasets are available in the supplemental material.

%% file: sec.feature_preservation.tex
\section{Feature Preservation Metrics}

Having defined the simplification techniques and generated their simplified outputs across a range of degrees, we now quantify how closely those outputs retain the visual features of the original data. We formalize a systematic way to estimate all the visual features described in~\Cref{sec:features} and define similarity metrics that measure preservation for each of them.

\subsection{Visual Feature Estimation}
All features in our study are estimated computationally from the time-series, without any human input. We distinguish between two classes of visual features: those with clear mathematical definitions and those that depend on user or data context (e.g., features such as trend, noise, spikes \& dips that depend more on user and context). This distinction arises because these features have multiple definitions in the visualization literature~\cite{proma2025visual}. Trend can be estimated by regression~\cite{correll2017regression} or the overall directional tendency~\cite{fan2022annotating}. We adopt the definitions and extraction procedures provided by Proma et al.~\cite{proma2025visual} for trend, periodicity, and noise using frequency domain analysis. 

Spikes \& dips are also context-dependent and rely on a user-specified threshold. We define them as points whose values deviate from the global mean by more than two standard deviations. We use this threshold as a simple criterion for identifying extreme deviations, which avoids false detections caused by noise while still capturing visually prominent outliers.

Regimes and change points also depend on contextual choices, such as the minimum segment length and the model that determines how many plateaus and transitions are considered meaningful. To obtain a data-driven estimate, we detect regime structure using the ruptures Python library~\cite{truong2020selective} with an $L^2$-norm based mean-shift model and a Bayesian Information Criterion (BIC) penalty. The model measures within-segment fit using $L^2$-norm, identifying change points where the 
segment-wise mean shifts sufficiently to reduce the overall reconstruction error. The BIC penalty prevents over-segmentation by charging a cost for each added segment: a split is kept only when it improves the model's fit by more than that cost, so segmentation stops once further splits are no longer worthwhile. We additionally impose a minimum segment length of 20 data points. This value was selected empirically. Across our datasets, segments shorter than 20 points consistently corresponded to noise-driven fluctuations rather than meaningful structural shifts. A minimum length of 20 prevents over-segmentation while still allowing detection of regime changes at moderate scales.

All remaining features are computed directly using their mathematical formulations described in~\Cref{sec:features}. We intentionally exclude \textit{anomalies} as a feature category. Anomalies are sometimes considered as spikes or dips~\cite{rosen2020linesmooth}, but the broader notion of an anomaly is inherently task- and domain-dependent and lacks a stable operational definition. Treating spikes and dips separately already captures the outlier-based definition of anomalies while avoiding the ambiguity of the broader concept. More generally, these features are detected independently rather than as mutually exclusive categories: a single time point may be a local extremum, a spike, and lie near a change-point at once. Because each feature is computed and compared with its own metric, we assign no precedence among them and measure their preservation separately.

\subsection{Similarity Metrics}
\label{sec_sim_metric}
To quantify feature-preservation errors, we compare each feature extracted from the original time-series with the corresponding feature extracted after simplification. Because different visual features differ in structure, some produce a full sequence of values (e.g., level, slope), others produce a single scalar (e.g., mean, roughness), and others produce sets of extrema (e.g., local extrema, spikes \& dips). A single similarity metric is not appropriate for all features. Therefore, we
use metric families that match the mathematical form of each feature.
 
\paragraph{Pointwise difference metrics}
Features such as level, slope, curvature, regression, trend, and noise yield a sequence of values. For these features, we compute pointwise error using the $L^1$ and $L^\infty$-norms.  $L^1$ measures the total pointwise deviation $\lnorm_{1}(X,Y)=\sum_{i=1}^{n} \left|x_i - y_i\right|$, and $L^\infty$ measures the largest deviations, $\lnorm_{\infty}(X,Y)=\max_i\left|x_i - y_i\right|$.

\paragraph{Scalar difference metrics}
Estimation of several visual features, such as mean, roughness, periodicity (amplitude and number of periods), and regimes \& change-points, yields a single scalar summary per output. For features such as mean and roughness, which are inherently scalar by definition (see~\Cref{sec:features}), preservation error is captured by the absolute difference
$\delta(F_1,F_2) = |F_1 - F_2|$.

Periodicity is likewise summarized using scalar quantities, namely the amplitude and the number of periods. We therefore compute preservation error using absolute differences in these two components.

Regimes and change-points are also reduced to scalar counts. We report preservation error using the difference in the number of regimes, defined as one plus the change-point count; differences in regime count therefore capture differences in the number of change-points. We do not compare the locations of individual change-points, as shifts in position or baseline are already reflected by features such as level, mean, and regression.

\paragraph{Topological distance metrics}
For features defined by extrema structure, such as local extrema and spikes \& dips, we use topological distances from persistent homology. We first identify local minima and maxima~\cite{rosen2020linesmooth, palshikar2009simple} and pair them using the persistence-based matching~\cite{suh2019topolines}. The resulting pairs form a persistence diagram~\cite{EdelsbrunnerHarer2010}, which encodes the prominence of each extremum.

To measure preservation error, we compare the persistence diagrams of the original and simplified charts using Wasserstein and Bottleneck distances. The Wasserstein distance provides a cumulative measure of structural discrepancy and is defined as $W_{1}(\overline{X},\overline{Y})=\inf_{\eta:\overline{X}\rightarrow\overline{Y}}\sum_{\overline{x}\in\overline{X}}\lVert \overline{x}-\eta(\overline{x}) \rVert_{1}$, where $\eta$ ranges over bijections between the two diagrams.
The Bottleneck distance captures the largest discrepancy in maxima–minima structure and is defined as $W_{\infty}(\overline{X},\overline{Y}) = \inf_{\eta: \overline{X} \rightarrow \overline{Y}} \sup_{\overline{x} \in \overline{X}} \left\| \overline{x}-\eta(\overline{x}) \right\|_\infty$.

%% file: sec.evaluation.tex
\section{Complexity Alignment and Evaluation}

Having estimated feature-preservation errors for each technique, our final step is to place all techniques on a common scale so their preservation can be compared directly. Comparing feature preservation across simplification techniques is challenging because algorithm parameters induce different degrees of visual complexity (see~\Cref{fig:simplification_examples}). We address this by aligning simplification outputs using a complexity-based measure and introducing a normalized scoring scheme that relates feature preservation to the degree of simplification.

\subsection{Quantifying Simplification}

Different techniques control simplification through different parameters such as window sizes, cutoff frequencies, target output sizes, or persistence thresholds. These parameter scales are not comparable across methods, and the same numerical change in a parameter may correspond to very different amounts of visual simplification. As a result, raw parameter values cannot serve as a unified simplification axis.

To enable comparison across simplification techniques, we require a measure that reflects the visual complexity of the resulting chart rather than the specific parameter used to produce it. To this end, we adopt Pixel Approximate Entropy (PAE)~\cite{ryan2018glance}. PAE assigns a scalar complexity value to a rendered line chart based on approximate entropy, capturing the degree of visual regularity and structure in the image.

By computing PAE for each of the 100 parameter levels per simplification technique, we obtain a mapping from technique-specific parameters to a shared complexity scale. This mapping allows us to compare different simplification techniques at comparable levels of visual simplification, even when their underlying parameters and mechanisms differ.

PAE depends on chart resolution, so we compute it using a fixed chart size of 1000 pixels wide by 375 pixels high. Fixing the resolution ensures that complexity scores are comparable across all simplification techniques and parameter settings. Because all techniques are rendered and evaluated at this same resolution, a change in resolution would rescale the PAE values but is not expected to substantially alter their relative ordering across techniques.

\begin{figure*}[t]
    \centering
    \captionsetup[subfloat]{labelfont=scriptsize,textfont=scriptsize}

    \includegraphics[width=\textwidth]{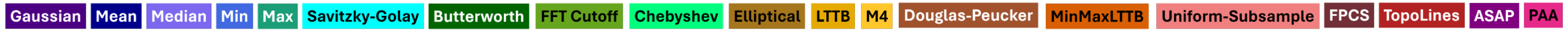}\\[6pt]

    \begin{minipage}[t]{0.325\textwidth}
        \centering
        \subfloat[{\scriptsize PAE vs.\ error metric}]{
            \includegraphics[width=\linewidth]{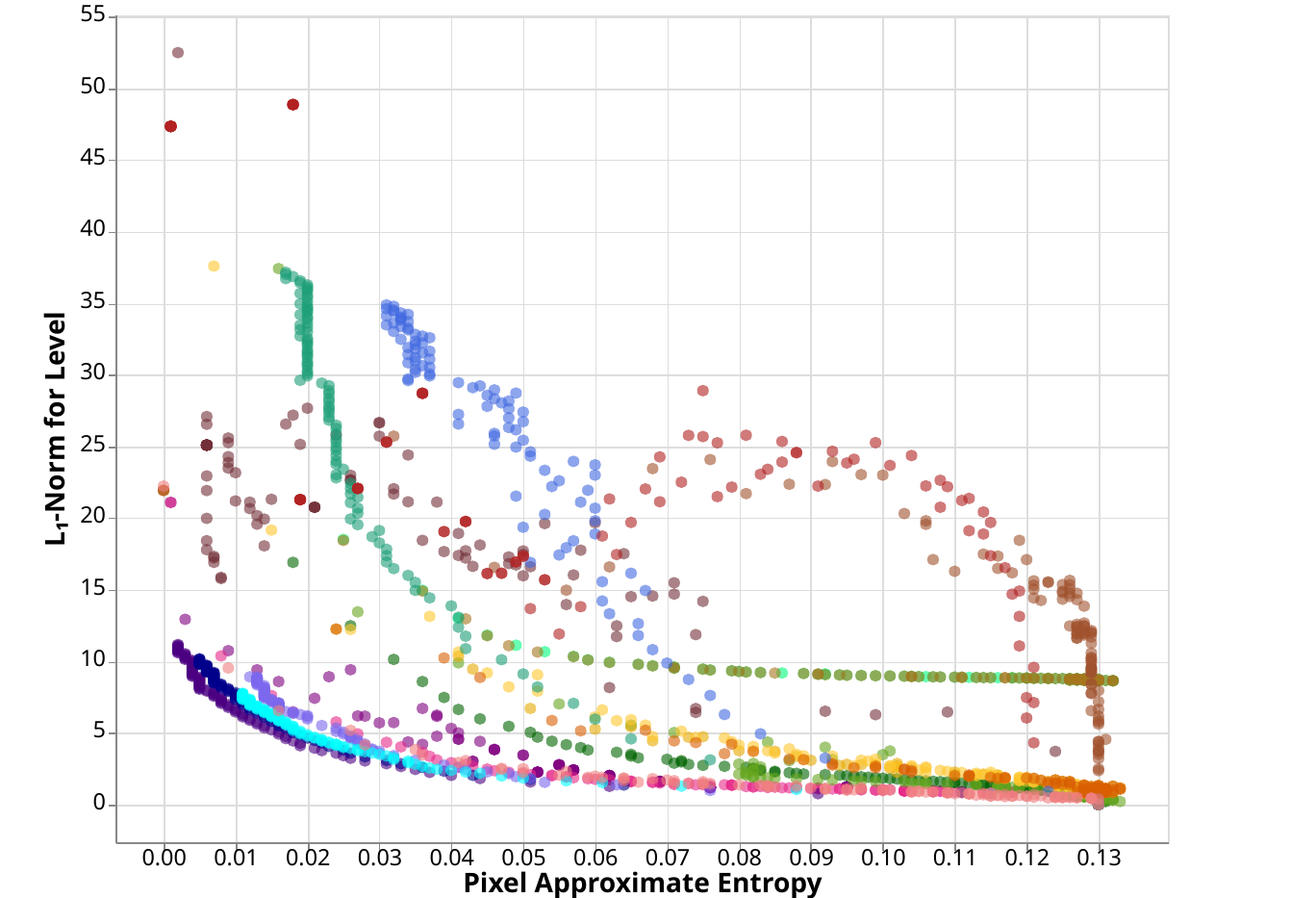}
            \label{fig:fc_raw}
        }
    \end{minipage}
    \hfill
    \begin{minipage}[t]{0.325\textwidth}
        \centering
        \subfloat[{\scriptsize Z-score normalization}]{
            \includegraphics[width=\linewidth]{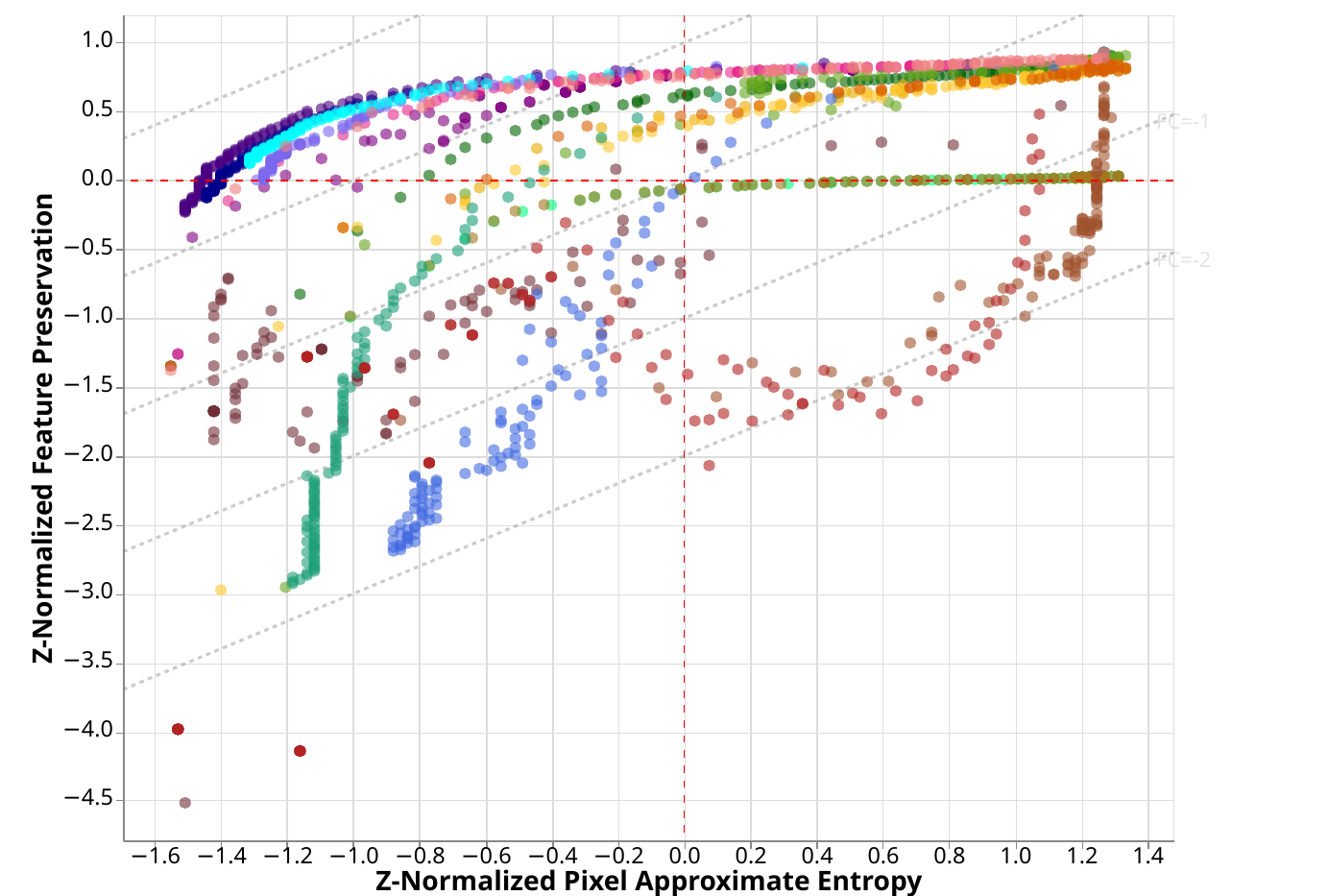}
            \label{fig:fc_zscore}
        }
    \end{minipage}
    \hfill
    \begin{minipage}[t]{0.325\textwidth}
        \centering
        \subfloat[{\scriptsize Quartile-band classification across simplification techniques}]{
            \includegraphics[width=\linewidth]{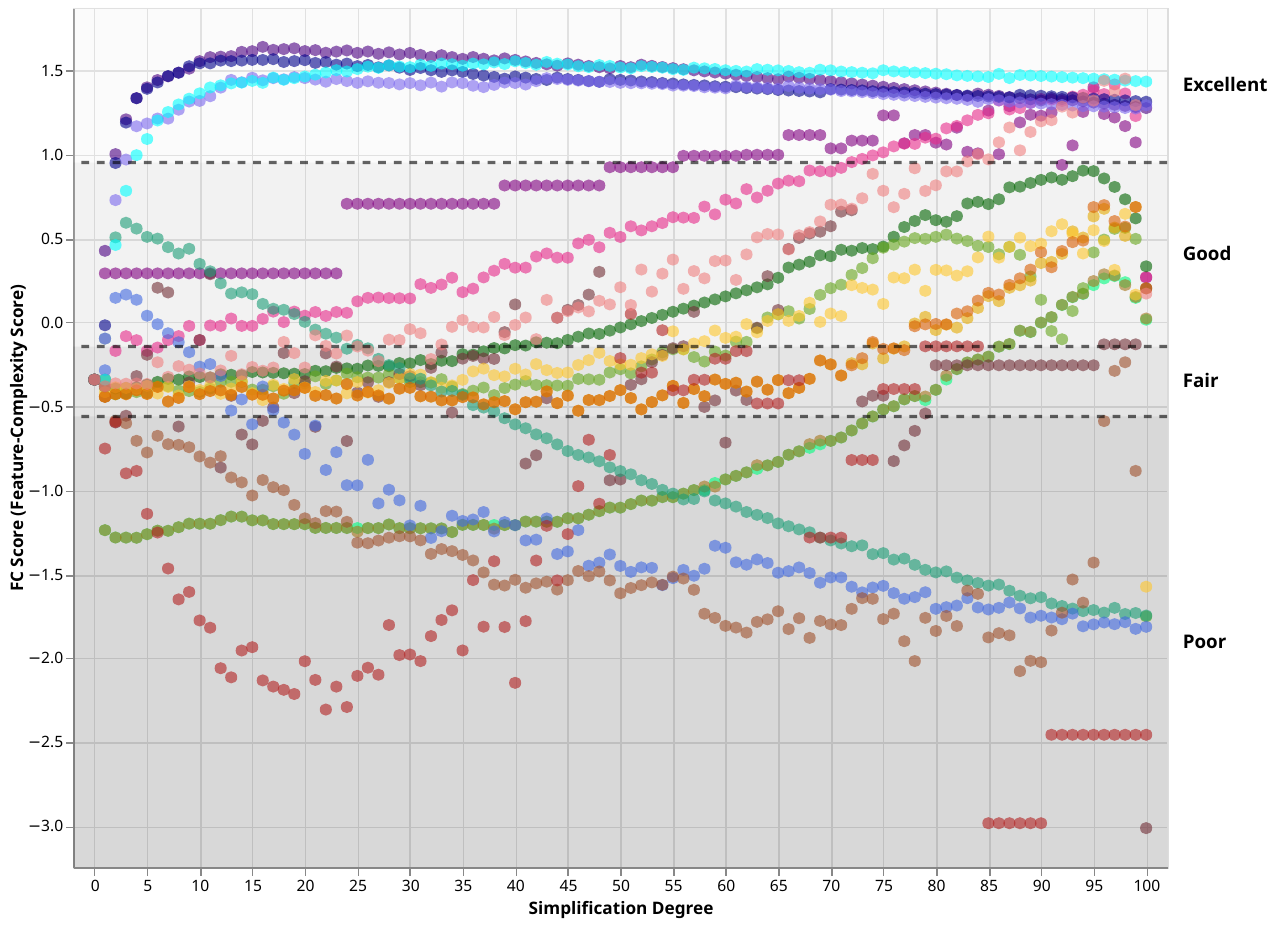}
            \label{fig:fc_band}
        }
    \end{minipage}
    \hfill

    \caption{
    Construction of the Feature-Complexity (FC) score on Apple stock price (\texttt{stock\_aapl\_price}) dataset for \textbf{Level} visual feature.
    (a)~PAE vs.\ feature–error values vary widely across simplification techniques, producing technique-specific PAE bands that
    prevent direct comparison.
    (b)~Z-score normalization aligns all configurations onto a shared visual-complexity axis, enabling
    cross-technique comparison.
    (c)~Quartile-band classification of simplification technique performances across simplification degrees: the top 25\% represent Excellent feature–complexity
    behavior, the next 25\% Good, the next 25\% Fair, and the bottom quartile Poor.
    }
    \label{fig:fc_score}
\end{figure*}

\subsection{FC Score: Normalized Feature--Complexity Alignment}
To compare how simplification techniques behave across different degrees of simplification, we need a way to place all configurations on a shared visual–complexity axis. However, each simplification technique produces a different scale and progression of PAE values. For example, the PAE at the 50th of 100 parameter levels for \methodGau may differ significantly from the PAE at the 50th of 100 parameter levels for \methodTDA. To address this, we introduce the Feature--Complexity (FC) score, which normalizes PAE across simplification techniques and relates it to feature preservation.

\Cref{fig:fc_raw} shows the relationship between PAE values and $L^1$-norm error for the \texttt{stock\_aapl\_price} dataset across all simplification techniques and degrees for the feature Level. In this format, PAE values cluster into bands specific to simplification techniques, and different methods span different complexity ranges, making cross-technique comparisons difficult.

To place all configurations on a common complexity axis, we apply z-score normalization to the PAE values for each dataset–feature pair. For each simplification technique and degree pair $i$, we compute
\[
z_i = \frac{\mathrm{PAE}_i - \mu_{\mathrm{PAE}}}{\sigma_{\mathrm{PAE}}},
\]
which yields a standardized measure of visual complexity that is comparable across simplification techniques.

The similarity metrics described in~\Cref{sec_sim_metric} produce feature-preservation errors, where larger values indicate greater deviation from the feature extracted from the original data. For each feature~$f$, we convert the error $E_{i,f}$ into a preservation score: $S_{i,f} = -E_{i,f}$,
because preservation and error are inversely related. We then apply z-score normalization across all these feature-preservation scores to get $S^{\,z}_{i,f}$.

We define the Feature-Complexity score for each simplification technique and degree pair $i$ as
\[
\text{FC}_i = S^{\,z}_{i,f} - z_i,
\]
so that higher values correspond to simultaneously higher feature preservation and lower normalized visual complexity. The subtraction reflects the intended role of FC as a trade-off measure rather than an absolute quality score. By normalizing both feature preservation and visual complexity, the difference captures how much a simplification configuration (combination of technique and simplification degree) preserves a feature relative to other configurations evaluated under the same dataset and feature. FC score thus captures the alignment between feature preservation and visual complexity within a given dataset and feature.

We initially summarized the performance of each simplification technique using mean FC scores for each feature. However, mean-based rankings can obscure meaningful differences when scores are numerically close. To address this limitation, we adopt a grading-based summary in this paper, which aggregates performance across different simplification degrees. Detailed ranking plots based on mean FC scores are provided in the supplemental material.

\subsection{Performance Grading}
\label{sec:grading}
While FC scores provide a continuous measure of feature preservation at different levels of simplification, their values vary substantially across features and datasets. To make performance comparable across these conditions, we convert the distribution of FC scores across 100 simplification degrees into a letter-grade system that reflects a simplification technique's ability to maintain feature quality across different simplification degrees. Rather than relying on a single best FC score, the grade captures the overall quality of a simplification technique using a GPA-style weighted average.

To evaluate the performance of a simplification technique in preserving a visual feature using a given similarity metric (e.g., worst-case trend preservation measured by $L^\infty$-norms for a Gaussian filter), we aggregate ratings across all datasets and simplification degrees. For each of the 80 datasets, the simplification technique is evaluated at 100 simplification degrees for a given feature, yielding 100 FC scores per dataset. Each FC score is assigned a rating (\emph{Excellent}, \emph{Good}, \emph{Fair}, or \emph{Poor}) based on its position within the metric-specific quartile distribution computed for that dataset (see~\Cref{fig:fc_band}): scores above the 75th percentile are labeled \emph{Excellent}, those above the 50th percentile \emph{Good}, those above the 25th percentile \emph{Fair}, and the remainder \emph{Poor}.

These ratings are then summarized using a weighted average, where \emph{Excellent} receives 4 points, \emph{Good} 3 points, \emph{Fair} 2 points, and \emph{Poor} 1 point. The resulting average score is converted to a letter grade using equal 0.6-point ranges: A for scores 3.4-4.0, B for 2.8-3.4, C for 2.2-2.8, D for 1.6-2.2, and F for scores below 1.6. This grade reflects the overall performance of the simplification technique in preserving the given feature across diverse datasets and simplification degrees.

To summarize, each step in this pipeline serves a specific role in making cross-technique comparison interpretable. We apply z-score normalization so that feature-preservation error and visual complexity, which are measured on different scales, can be combined without one dominating by magnitude. We subtract normalized complexity from normalized preservation so that the FC score expresses the fidelity--complexity trade-off as a single value rather than an absolute quality score. We then summarize the resulting scores using quartile-based ratings and a GPA-style grade, which capture performance across 100 simplification degrees and 80 datasets while avoiding the sensitivity of mean-based rankings to numerically close scores.

%% file: sec_results.tex
\section{Results}

Based on the grading framework described in \Cref{sec:grading}, we analyze how simplification techniques perform across datasets and feature types. Grades summarize feature-preservation quality across all simplification degrees for each technique. For each visual feature, we compute a letter grade reflecting the technique's performance by identifying the \emph{mode grade}, the most frequently occurring grade across all 80 datasets. This mode grade characterizes the typical feature-preservation quality a technique achieves.

To capture how consistently a technique exhibits this typical behavior across datasets, we compute a \emph{deviation}, defined as the mean absolute distance between each dataset's grade and the mode grade. Low deviation indicates stable, predictable behavior across datasets, while high deviation indicates strong sensitivity to data characteristics.

Using mode grades to summarize typical feature preservation quality and deviation to assess stability across datasets, we examine performance variation across visual features and the extent to which feature preservation behavior depends on the underlying data. Together, these analyses provide a structured view of which simplification techniques are effective for preserving certain visual features and how reliably these techniques behave across diverse datasets.

\subsection{Grades Across Visual Features}
Determining mode grades across datasets reveals that variation in feature-preservation performance is driven primarily by the choice of simplification technique rather than by the visual feature itself. In ~\Cref{fig:teaser}, the vertical patterns (variation across simplification techniques for a single feature metric) show shifts from A to F grades, while horizontal patterns (variation across features for a single technique) remain relatively stable, indicating broadly uniform performance rather than feature-specific behavior.

Several techniques demonstrate consistently high or low grades across nearly all visual feature metrics. For example, transformers such as \methodGau{} (21/21 A grades), \methodMean{} (20 A, 1 C), and \methodMed{} (20 A, 1 C) receive A grades across nearly all visual feature metrics, indicating consistently high feature-preservation performance across features. In contrast, reducers such as \methodDP{} (14 F, 7 D) and \methodMFour{} (14 D, 6 F, 1 C) receive low grades on most metrics. In these cases, preservation quality varies little across feature types and is instead dominated by the choice of simplification technique.

In contrast, a smaller subset of techniques exhibits clear feature-dependent behavior. Extreme-value filters, such as \methodMin{} and \methodMax{}, show higher grades on extrema-related metrics but lower grades on features like level, trend, and mean. Similarly, \methodFPCS{} exhibits uneven performance across features, achieving higher grades in mean, slope, curvature, and noise preservation, while performing poorly in level, regimes, extrema, trend, and roughness preservation. Aggregators such as \methodASAP{} and \methodPAA{} occupy an intermediate position, achieving mostly B- and C-level grades, with some variation across features. They also consistently outperformed reducers, but fell short of the uniformly high-performing transformers. These patterns suggest that certain techniques interact more selectively with visual features, resulting in pronounced horizontal variation in the grade matrix. 

Overall, the grade distributions suggest that differences in feature preservation are primarily driven by algorithmic behavior rather than by variations in feature difficulty. While certain techniques exhibit feature-specific strengths or weaknesses, most simplification methods tend to preserve or distort visual features in a relatively uniform manner across the feature set.

\subsection{Consistency Across Visual Features}
While mode grades summarize the most common performance level a simplification technique achieves in preserving visual features across datasets, they do not indicate how consistently it behaves. To capture this dimension, we compute the \emph{deviation}, which measures how far a technique's grades fall from the mode grade across datasets.

We map each letter grade to a numeric value (A as 4, B as 3, C as 2, D as 1, F as 0), so that adjacent grades differ by one and the distance between two grades reflects how many grade levels apart they are. Formally, for a given visual feature, let $M$ be the numeric value of the mode grade and $g_i$ be the numeric grade a technique receives on dataset $i$. Across $n$ datasets, the deviation is the mean absolute distance from the mode:

$$\text{Deviation} = \frac{1}{n}\sum_{i=1}^{n} \left| g_i - M \right|$$

This deviation reflects the \emph{data dependency} of each method, and is expressed in grade units: zero means every dataset received the mode grade, while larger values mean grades fall further from it, either because more datasets differ or because the differences are larger. We classify deviation at natural fractions of a grade:

\begin{itemize}
    \item \textbf{Consistent (Deviation $<0.25$):} Grades almost always match the mode, with only occasional single-grade differences, indicating that the technique behaves similarly regardless of data characteristics.

    \item \textbf{Moderately data-dependent (Deviation $0.25$-$0.5$):} Grades differ from the mode on a noticeable share of datasets, usually by a single grade, suggesting partial dependency on dataset structure.
    
    \item \textbf{Highly data-dependent (Deviation $>0.5$):} Grades differ from the mode on a large share of datasets, and a notable portion fall two or more grades away. This reveals a strong interaction between the simplification technique and data characteristics.
\end{itemize}

When two grades occur with equal frequency, both are considered mode grades. In such cases, we measure each dataset's distance from the midpoint of the two tied grades rather than from either grade alone, since a technique that does not settle on a single typical grade is itself exhibiting variation. 

The deviation is indicated in~\Cref{fig:teaser} across all 19 simplification techniques and 21 feature preservation metrics, with varying orange-colored lines under the grade histogram. The observed deviations span a wide range, from consistency to high variability. We classify each technique by the deviation category into which the largest share of its feature metrics fall. Several systematic patterns emerged from our analysis:
\begin{itemize}

    \item \textbf{Predominantly data-independent:} Six simplification techniques exhibit predominantly low deviation (below 0.25) across most feature metrics, indicating that their behavior is largely predictable across diverse datasets. \methodGau{} demonstrates the highest overall stability with low deviation for 16 of 21 metrics, followed by \methodMean{} (13/21). Four further techniques retain a low-deviation plurality while being more sensitive to dataset characteristics: \methodSG{} (10/21), \methodBW{} (10/21), \methodMed{} (9/21), and \methodDP{} (8/21). 
    
    Across these methods, extrema and spikes \& dips preservation frequently exhibit higher deviation, indicating that preservation of extreme structures is inherently more data dependent even for otherwise stable techniques. \methodGau{} reflects this pattern, receiving a modal A grade for extrema and spikes \& dips while still showing elevated deviation for these features, as shown in~\Cref{fig:teaser}.

    \item \textbf{Moderately data-dependent methods:} For seven simplification techniques, the largest share of feature metrics falls into the moderate deviation category: \methodCheb{} (15/21 metrics with medium deviation), \methodEllip{} (15/21), \methodMFour{} (13/21), \methodPAA{} (12/21), \methodUni{} (10/21), \methodLTTB{} (9/21), and \methodMinMaxLTTB{} (8/21). 
    
    For these techniques, moderate deviation across a broad set of metrics indicates that preservation quality depends on dataset characteristics without collapsing into either consistent stability or extreme variability. This places them in an intermediate position of dataset sensitivity, more data-dependent than the predominantly stable methods but more predictable than the highly data-dependent ones.

    \item \textbf{Mostly data-dependent methods:} Six methods exhibit substantial high deviation: \methodMin{} (20/21 metrics with high deviation), \methodFPCS{} (19/21), \methodASAP{} (17/21), \methodTDA{} (15/21), \methodCut{} (12/21), and \methodMax{} (10/21). \methodMin{}, \methodTDA{}, and \methodASAP{} have no low-deviation metrics across the 21 features, indicating an absence of stable preservation across feature types. \methodMin{} and \methodFPCS{} are the most extreme cases, with high deviation on nearly all metrics, indicating that few features are preserved predictably across datasets. 
    
    For these methods, preservation quality depends on whether the dataset aligns with the assumptions implicit in the technique. For example, \methodASAP{} relies on autocorrelation-based heuristics to adaptively select aggregation windows, yielding strong preservation when the data has clear structure but more variable outcomes on irregular time-series. Similar patterns appear for other adaptive or extreme-value-based techniques.
\end{itemize}

A simplification technique with a high mode grade but high deviation (e.g., \methodMed{} on features such as extrema or curvature) may be excellent for specific dataset characteristics but unreliable elsewhere. Conversely, a technique with a low mode grade and low deviation (e.g., \methodDP{}, which consistently receives lower grades) is unsuitable in most cases.

%% file: sec.discussion.tex
\section{Discussion}
\label{sec:discussion}
Our results show that typical feature-preservation performance is driven primarily by the choice of simplification technique. In the grade matrix (\Cref{fig:teaser}), grades vary sharply across techniques for the same feature metric (strong vertical variation), while many techniques exhibit relatively stable grades across the feature set (weaker horizontal variation). This pattern indicates that, in our grading summaries, differences in preservation are more strongly associated with technique behavior than with a single feature being uniformly easy or hard to preserve.

At the same time, the deviation analysis shows that feature preservation is not equally stable across datasets. Several techniques achieve low deviation for a large subset of metrics (e.g., \methodGau{}, \methodMean{}, \methodSG{}, \methodDP{}, \methodBW{}, \methodMed{}), whereas others exhibit high deviation across a large share of metrics (e.g., \methodFPCS{}, \methodASAP{}, \methodCut, and the extreme-value filters \methodMin{}, \methodMax{}, and \methodTDA{}). Even among stable techniques, deviation increases for metrics tied to extrema and spikes \& dips, indicating that preservation of these feature categories is more dataset-sensitive. Based on these findings, we distill the following main takeaways.

\paragraph{\textbf{Transformers preserve features reliably but do not reduce data density}}
Across datasets and visual features, transformers such as \methodGau, \methodMean, and \methodMed frequently receive higher mode grades than other methods, indicating strong feature preservation. Because transformers modify values while retaining all sample positions, they avoid the structural loss introduced by point removal and rarely receive consistent lower grades across visual feature preservation.

Importantly, all transformers preserve the original data density. While they reduce visual complexity through value modification, they do not reduce data density. As a result, transformers are best suited to scenarios where preserving fine-grained structure is prioritized; however, they are not appropriate when reducing the number of rendered data points is the primary goal.

\paragraph{\textbf{Reducers trade fidelity for data point reduction, while aggregators mitigate, but do not eliminate this trade-off}}
Reducers necessarily discard samples to achieve compression, and this structural loss is reflected in consistently lower grades compared to transformers. This effect is particularly pronounced for \methodDP{}, which consistently yields low grades across most visual features. Even among higher-performing reducers such as \methodFPCS{} (11 A/B grades, 2 F grades), feature preservation remains constrained by the fundamental limitation that removed points cannot contribute to retained structure. These grades measure the price of compression, a capability that transformers lack entirely.

Aggregators reduce resolution but differ in how they do so. By combining transformation and point reduction, methods such as \methodASAP{} (11 A/B grades, 0 F grades) and \methodPAA{} (6 B grades, 0 F grades) retain visual features more effectively than most reducers. While aggregators do not match the feature preservation of high-fidelity transformers such as \methodGau{} and \methodMean{}, they occupy a middle ground between full-resolution preservation and direct geometric reduction, making them appropriate when both complexity minimization and point reduction are required.

\paragraph{\textbf{The taxonomy exposes aggregation as a compositional design space}}
Our taxonomy makes explicit that aggregation combines two operations: a transformer that shapes the signal and a reducer that selects representative samples. This perspective helps explain why aggregators received higher grades than reducers for preserving many visual features, while still falling short of high-fidelity transformers.

More broadly, the taxonomy suggests a design space in which transformers and reducers can be combined intentionally. While we do not evaluate arbitrary combinations in this work, the results suggest that separating these stages provides a principled approach to reasoning about fidelity--compression trade-offs and motivates future exploration of feature-aware simplification techniques.

\paragraph{\textbf{Consistency vs.\ data dependence across datasets}}
Mode grades summarize the typical level of feature preservation a simplification technique achieves, but they do not indicate how stable that performance is across datasets. Our deviation analysis reveals that consistency varies substantially across techniques, highlighting when algorithm behavior is predictable and when it is strongly modulated by dataset structure.

At a high level, certain techniques exhibit stable and consistently strong behavior. In particular, \methodGau{} achieves uniformly high mode grades across most datasets and visual features, accompanied by low deviation for a large fraction of metrics. This indicates that \methodGau{} is not only a strong performer on average, but also a reliable one across diverse data domains. In contrast, \methodDP{} consistently receives low mode grades across many features, accompanied by low deviation. This reflects generally poor feature preservation. However, this characterization is not absolute. When examining dataset-level mode grades (see supplemental material), \methodDP{} achieved B-level mode grades on multiple \texttt{climate\_prcp} datasets, demonstrating that even techniques with poor overall performance can exhibit competitive behavior on specific data structures.

This pattern underscores that feature-preservation behavior is often data-dependent rather than universally good or bad. Deviation analysis makes this explicit. Techniques such as \methodMean{}, \methodSG{}, and \methodMed{} exhibit relatively low deviation across many features, indicating predictable behavior across datasets. In contrast, methods such as \methodMin{}, \methodMax{}, \methodTDA{}, \methodCut{}, \methodFPCS{}, and \methodASAP{} exhibit high deviation across most features, with grades ranging from A to F, depending on the dataset. For these techniques, performance cannot be inferred from a single representative dataset or mode grade.

Taken together, these results show that while some techniques (e.g., \methodGau{}) are consistently strong choices and others (e.g., \methodDP{}) are typically weaker, feature preservation cannot be characterized by global labels alone. Mode grades identify typical behavior, while deviation reveals the extent to which that behavior generalizes across datasets. Both are necessary to distinguish techniques that are reliably good, reliably poor, or highly data dependent.

\paragraph{\textbf{No feature is inherently hard to preserve, but extrema and spikes \& dips are strongly data-dependent}}
Deviation analysis reveals that preservation of extrema and spikes \& dips is consistently more data-dependent than other visual features. As shown in \Cref{fig:teaser}, for both extrema and spike \& dips related metrics, the majority of the simplification techniques exhibit either moderate or high deviation, indicating that no technique preserves these features in a stable manner across datasets. However, on \texttt{climate\_lax\_tmax}, which has prominent spikes among a clear periodic pattern, most techniques received A-level grades for preserving spikes \& dips, whereas on the mostly smooth \texttt{unemployment\_info} none did. This indicates the importance of considering the structure of the data when choosing a technique to preserve spikes \& dips.

At the same time, examination of the grade distributions does not reveal any visual feature that is uniformly easy or hard to preserve. For example, high-fidelity transformers such as \methodGau{} receive A-level grades across all feature metrics, including extrema and spikes \& dips, indicating that these features are not inherently difficult to preserve in absolute terms. Rather, the elevated deviation observed for extrema and spikes \& dips related metrics reflects sensitivity to dataset characteristics rather than a global limitation of the feature itself.

\paragraph{\textbf{Feature preservation is context-dependent, not universally desirable}}
Although several techniques achieve uniformly high grades across all features, strong preservation is not inherently beneficial in all contexts. For example, high-fidelity transformers such as \methodGau{} receive A-level grades for noise and roughness preservation across datasets, reflecting their tendency to retain high-frequency variation (see~\Cref{fig:teaser}). However, in many analytical settings, preserving noise or roughness is not a goal, and may instead obscure higher-level structure.

The grade distributions in \Cref{fig:teaser} show that other techniques achieve different trade-offs between fine-scale variation and higher-level features. For instance, \methodFPCS{} attains A-level grades for slope-related metrics while receiving B-level grades for noise preservation, indicating that moderate noise attenuation can coincide with stronger preservation of structural features on certain datasets.
These patterns highlight that the quality of feature preservation must be interpreted relative to the analytic intent. Uniformly high grades across all features do not imply universal suitability, and techniques that moderately attenuate noise may be preferable when the goal is to emphasize trend or slope rather than fine-scale variation.

%% file: sec.conclusion.tex
\section{Conclusion}
Line chart simplification is a nuanced process in which reducing visual complexity must be carefully managed to preserve the visual features that support interpretation and understanding. In this work, we introduce a visual feature-based taxonomy of line chart simplification, along with a systematic evaluation framework that characterizes how commonly used algorithms preserve or distort these features across datasets and simplification levels. Rather than assessing simplification strategies through task performance, our taxonomy makes explicit which visual structures each algorithm supports, sacrifices, or distorts.

A central contribution of this paper is to enable feature-driven reasoning about simplification techniques by articulating their underlying algorithmic principles. Rather than assuming that methods within a category behave similarly, we evaluate feature preservation at the level of individual techniques and identify recurring patterns that align with how those techniques operate on the signal. These patterns motivate a taxonomy organized around three principles: transformation, reduction, and aggregation, that link how an algorithm modifies data to how it preserves visual features. Beyond grouping methods, this link supports reasoning about techniques not explicitly evaluated and guides the design of new or hybrid methods: combining a transformation that stabilizes certain features with a reduction strategy that controls point count can be expected to inherit the feature-preservation properties of each. The taxonomy thus serves both to characterize existing techniques and to anticipate the behavior of future ones.

To ensure tractability and comparability, our evaluation makes several deliberate design choices that bound the scope of our analysis. We vary a single primary simplification parameter per algorithm, while holding secondary parameters fixed using the recommended defaults from prior work~\cite{rosen2020linesmooth}. Although this captures the dominant effect of simplification for most methods, alternative parameter settings could yield different trade-offs in feature preservation. While our normalization mitigates differences in absolute complexity scales, a more exhaustive exploration of multi-parameter configurations remains an open direction. Several contextual features, such as spikes \& dips, regimes, and change-points, also rely on threshold or model-selection parameters that we selected and justified empirically. A systematic study of their sensitivity across all techniques, degrees of simplification, and datasets is outside the scope of this study; therefore, results for these features should be read with these choices in mind, and we leave such a study to future work. Our evaluation is also computational, measuring feature preservation and visual complexity directly from the data; how viewers perceive these features may differ from what our metrics capture. Extending the taxonomy to richer parameter interactions and validating our measures against human judgments are both important directions for future work.

In summary, this paper presents a unifying framework for understanding, evaluating, and selecting line chart simplification strategies from the perspective of visual feature preservation. By grounding algorithm behavior in a shared taxonomy and systematic evidence, we support more transparent, justifiable, and feature-aware simplification decisions.